\documentclass[10pt]{article}
\usepackage{amssymb}
\usepackage{amsmath}
\usepackage{mathrsfs}
\usepackage{graphicx}
\usepackage{fancyhdr}
\usepackage{cite}
\usepackage{mathrsfs} 
\usepackage{subeqnarray} 
\usepackage{subfigure} 

\newcommand{\strutl}{\vrule width 0pt height 0pt depth 16pt}

\newcommand{\goodgap}{
\hspace{\subfigtopskip}%
\hspace{\subfigbottomskip}}%
 
\begin{document}

\huge

\begin{center}
Analytical and numerical expressions for the number of atomic configurations 
contained in a supershell\end{center}

\vspace{0.5cm}

\large

\begin{center}
Jean-Christophe Pain$^{a,}$\footnote{jean-christophe.pain@cea.fr} and Michel Poirier$^b$
\end{center}

\normalsize

\begin{center}
$^a$CEA, DAM, DIF, F-91297 Arpajon, France\\

$^b$Universit\'e Paris-Saclay, CEA, CNRS, LIDYL, F-91191 Gif-sur-Yvette, France
\end{center}

\vspace{0.5cm}


\begin{abstract}
We present three explicit formulas for the number of electronic configurations in an atom, i.e. the number of ways to distribute $Q$ electrons in $N$ subshells of respective degeneracies $g_1$, $g_2$, ..., $g_N$. The new expressions are obtained using the generating-function formalism. The first one contains sums involving multinomial coefficients. The second one relies on the idea of gathering subshells having the same degeneracy. A third one also collects subshells with the same degeneracy and leads to the definition of a two-variable generating function, allowing the derivation of recursion relations. All these formulas can be expressed as summations of products of binomial coefficients. 
Concerning the distribution of population on $N$ distinct subshells of a given 
degeneracy $g$, analytical expressions for the first moments of this distribution 
are given. The general case of subshells with any degeneracy is analyzed through 
the computation of cumulants. A fairly simple expression for the cumulants at any 
order is provided, as well as the cumulant generating function. Using Gram-%
Charlier expansion, simple approximations of the analyzed distribution in terms 
of a normal distribution multiplied by a sum of Hermite polynomials are given. 
These Gram-Charlier expansions are tested at various orders and for various 
examples of supershells. When few terms are kept they are shown to provide simple 
and efficient approximations of the distribution, even for moderate values of the 
number of subshells, though such expansions diverge when higher order terms are 
accounted for. The Edgeworth expansion has also been tested. Its accuracy is 
equivalent to the Gram-Charlier accuracy when few terms are kept, but it is much 
more rapidly divergent when the truncation order increases. While this analysis 
is illustrated by examples in atomic supershells it also applies to more general 
combinatorial problems such as fermion distributions.
\end{abstract}

\section{Introduction}

The knowledge of the number of atomic configurations (i.e. the number of possible ways to distribute $Q$ electrons in $N$ subshells of respective degeneracies $g_1$, $g_2$, ..., $g_N$) is important for the computation of atomic structure and spectra \cite{BAUCHE87,BAR89,XU06,GILLERON09,PAIN12,BAUCHE15} and is a fundamental problem of statistical physics \cite{PRATT00,PONOMARENKO06,KOSOV08,CHAKRABORTY15}. However, it is a difficult combinatorial problem (belonging to the class of the so-called ``bounded partitions'' \cite{ANDREWS76,WIRSCHING98,GLUCK13}) and the number of electronic configurations is usually evaluated numerically by direct  multiple summations requiring the computation of nested-loops. A few years ago, efficient double recursion relations, on the number of electrons and the number of orbitals, were published \cite{GILLERON04,WILSON07,PAIN09}. However, we could not find in the literature an analytical expression valid in any case. 
For this reason, in this paper we develop various analytical and numerical 
methods providing this number of configurations. As part of the above quoted 
bibliography suggests, the present analysis is not limited to the number of 
configurations obtained by distributing $Q$ electrons in a list of subshells, 
but deals with more general combinatorial questions related, e.g., to fermion 
statistics.

The generating function for the number of configurations is introduced in section \ref{sec1}, along with some of its interesting properties. The first expressions involving multinomial coefficients is presented in section \ref{sec2}, and the second expression, obtained by partitioning the subshells into iso-degeneracy groups, is derived in section \ref{sec3}. 
Focusing on the case of susbshells with the same degeneracy, a two-variable 
generating function allows us to obtain several recurrence relations (section 
\ref{sec:rec_sameg}) and to compute moments at any order (section 
\ref{sec:moments}). Furthermore, the cumulants of this distribution as well 
as the cumulant generating function are obtained analytically in section 
\ref{sec:cumulants}. The availability of these cumulants allows us to derive 
simple approximations for this number of configurations using a Gram-Charlier 
and Edgeworth expansion in sections \ref{sec:Gram-Charlier} and 
\ref{sec:Gram-Charlier} respectively. Concluding remarks are finally given.

\section{Generating function of the problem}\label{sec1}

We have to find the number of integer solutions of $Q=q_1+q_2+q_3+...$ with the restrictions : $0\leq q_1\leq g_1$, ..., $0\leq q_N\leq g_N$. 
Such constraints can be efficiently accounted for using generating functions\cite{KATRIEL83,KATRIEL89}.
This number of solutions being denoted $\mathcal{C}\left(Q,N\right)$, we define 
the generating function with

\begin{subequations}\begin{align}
G(x,N)=&\sum_{Q=0}^{\infty}x^{Q}\mathcal{C}\left(Q,N\right)\\
=&\sum_{Q=0}^{\infty}x^Q\sum_{\left\{q_1, q_2, \cdots, q_N\right\}}\delta_{Q,q_1+q_2+\cdots+q_N} 
 \theta\left(g_1-q_1\right)\cdots\theta\left(g_N-q_N\right),
\end{align}\end{subequations}
where $\delta$ represents the Kronecker symbol and $\theta$ the Heaviside function. One gets

\begin{equation}
G\left(x,Q\right)=\sum_{\left\{q_i\right\}}x^{q_1+q_2+\cdots+q_N}\theta\left(g_1-q_1\right)\cdots\theta\left(g_N-q_N\right).
\end{equation}

\noindent Since the quantities $q_i$ are independent, one has

\begin{equation}
G\left(x,Q\right)=\sum_{q_1=0}^{g_1}x^{q_1}\cdots\sum_{q_N=0}^{g_N}x^{q_N},
\end{equation}
i.e.,

\begin{equation}
G(x,N)=\prod_{i=1}^N\left[\frac{1-x^{g_i+1}}{1-x}\right]
 =\frac{1}{(1-x)^N}\prod_{i=1}^N\left(1-x^{g_i+1}\right),
\end{equation}
with

\begin{equation}\label{rel1}
\frac{1}{(1-x)^N}=\sum_{i=0}^{\infty}\binom{N-1+i}{N-1}x^i.
\end{equation}

\noindent If all the orbitals had the same degeneracy, we would have

\begin{equation}\label{rel2}
\prod_{i=1}^N\left(1-x^{g+1}\right)=\left(1-x^{g+1}\right)^N=\sum_{k=0}^N(-1)^k\binom{N}{k}x^{k(g+1)}
\end{equation}
and, combining Eqs. (\ref{rel1}) and (\ref{rel2})

\begin{equation}
\mathcal{C}(Q,N)=\sum_{k=0}^{\lfloor N/(g+1)\rfloor}\binom{N}{k}(-1)^k
 \binom{N-1+Q-k(g+1)}{N-1},\label{eq:CQN_one_g}
\end{equation}
where $\lfloor x\rfloor$ denotes the integer part of $x$. However, since all the orbitals do not in general have the same degeneracy, the problem is more complicated. Let us take the example of four orbitals with degeneracy $g_1,g_2,g_3,g_4$. In the present case, this generating function involves the product

\begin{equation}\begin{split}
\prod_{i=1}^4\left(1-x^{g_i+1}\right)=&1-x^{g_1+1}-x^{g_2+1}-x^{g_3+1}-x^{g_4+1}
 +x^{g_1+g_2+2}+x^{g_1+g_3+2}+x^{g_1+g_4+2}\\
& +x^{g_2+g_3+2}+x^{g_2+g_4+2}+x^{g_3+g_4+2}
 -x^{g_1+g_2+g_3+3}-x^{g_1+g_2+g_4+3}\\
& -x^{g_1+g_3+g_4+3}-x^{g_2+g_3+g_4+3}
 +x^{g_1+g_2+g_3+g_4+4},
\end{split}\end{equation}
which can be expressed in terms of the so-called symmetric functions \cite{BALANTEKIN01,PAIN11}. Knowing the generating function, one can now write $\mathcal{C}(Q,N)$ as a contour integral

\begin{subequations}\begin{align}
\mathcal{C}\left(Q,N\right)=&\frac{1}{2i\pi}\oint\frac{dz}{z^{Q+1}}G\left(z,N\right)\\
=&\frac{1}{2i\pi}\oint\frac{dz}{z^{Q+1}}\prod_{i=1}^N\left[\frac{1-z^{g_i+1}}{1-z}\right].
\end{align}\end{subequations}

\noindent Assuming that the number of electrons $Q$ and the number of orbitals $N$ are large, one finds (following the asymptotics of partitions of Hardy-Ramanujan \cite{ANDREWS76})

\begin{equation}
\mathcal{C}(Q,N)=\frac{1}{2i\pi}\oint\frac{dz}{z}e^{S_{N,Q}(z)},
\end{equation}
with 

\begin{equation}
S_{N,Q}(z)=\sum_{i=1}^N\ln\left(\frac{1-z^{g_i+1}}{1-z}\right)-Q\ln z,
\end{equation}
and one has to find $z_0$ such that $\left.\frac{dS_{N,Q}}{dz}\right|_{z_0}=0$. However, it is difficult to find some large quantities in the present case. Therefore, we usually make the calculation using a recursion relation \cite{PAIN02}

\begin{align}\nonumber
\mathcal{C}\left(Q,N\right)=&\sum_{i=0}^Q\mathcal{C}(Q-i,N-1)\theta(g_N-i)\\
 =&\sum_{i=0}^{\min(Q,g_N)}\mathcal{C}(Q-i,N-1),
\label{eq:recurnbconfig}\end{align}
where $g_N$ is the last-orbital degeneracy. The recurrence is 
initialized by $\mathcal{C}\left(Q,0\right)=\delta_{Q,0}$.

One may note that, in a different context, formula (\ref{eq:CQN_one_g}) has been used by 
Crance (see Appendix in Ref. \cite{Crance84}) to calculate the proportion of neutral atoms 
in a statistical description of multiple ionization.

\section{First exact expression involving multinomial coefficient}\label{sec2}

The number of atomic configurations of $Q$ electrons in $N$ subshells is related to the generating function $G(x,N)$ by

\begin{equation}
\mathcal{C}\left(Q,N\right)=\left.\frac{1}{Q!}\frac{\partial^Q}{\partial x^Q}G(x,N)\right|_{x=0}.
\end{equation}

\noindent The recursion relation (\ref{eq:recurnbconfig}) can be obtained from this relation. Using the Leibniz rule for the derivative of a product of two functions, we obtain

\begin{equation}
\mathcal{C}\left(Q,N\right)=\left.\frac{1}{Q!}\sum_{i=0}^Q\binom{Q}{i}
 \frac{\partial^i}{\partial x^i}\frac{1}{(1-x)^N}\right|_{x=0}
 \left.\frac{\partial^{Q-i}}{\partial x^{Q-i}}\prod_{i=1}^N\left(1-x^{g_i+1}\right)\right|_{x=0}.
\end{equation}

\noindent We have

\begin{equation}
\left.\frac{\partial^i}{\partial x^i}\frac{1}{(1-x)^N}\right|_{x=0}=i!\binom{i+N-1}{i}
\end{equation}
and

\begin{equation}
 \left.\frac{\partial^{Q-i}}{\partial x^{Q-i}}\prod_{i=1}^N\left(1-x^{g_i+1}\right)\right|_{x=0}=
 \sum_{\vec{\alpha}/\sum_{j=1}^N\alpha_j=Q-i}\frac{(Q-i)!}{\alpha_1!\alpha_2!\alpha_3!...\alpha_N!}
 \prod_{j=1}^N\frac{\partial^{\alpha_j}}{\partial x^{\alpha_j}}\left.\left(1-x^{g_j+1}\right)\right|_{x=0},
\end{equation}
where $\vec{\alpha}=\left(\alpha_1,\alpha_2,\cdots,\alpha_N\right)$. The quantity

\begin{equation}\label{eq:multinomial}
\binom{Q-i}{\alpha_1,\alpha_2\cdots,\alpha_N} = 
\frac{(Q-i)!}{\alpha_1!\alpha_2!\cdots\alpha_N!}
\end{equation}
is the multinomial coefficient. 
It can be expressed in numerous ways, including a product of binomial coefficients 

\begin{equation}
\binom{Q-i}{\alpha_1,\alpha_2\cdots,\alpha_N}= \delta_{Q-i,\alpha_1+\cdots\alpha_N}
 \binom{\alpha_1}{\alpha_1}\binom{\alpha_1+\alpha_2}{\alpha_1}\cdots\binom{Q-i}{\alpha_N}.
\end{equation}

\noindent We have also, if $\alpha_j\ne 0$

\begin{equation}
\frac{\partial^{\alpha_j}}{\partial x^{\alpha_j}}\left.\left(1-x^{g_j+1}\right)\right|_{x=0}
 =-(g_j+1)!\times\delta_{\alpha_j,g_j+1}
\end{equation}

\noindent and we get finally

\begin{equation}
\mathcal{C}(Q,N)=\frac{1}{Q!}\sum_{i=0}^Qi!\binom{Q}{i}\binom{i+N-1}{i}
 \sum_{\vec{\alpha}/\sum_{j=1}^N\alpha_j=Q-i}\frac{(Q-i)!}{\alpha_1!\alpha_2!...\alpha_N!}
 \prod_{j=1}^N\left(\delta_{\alpha_j,0}-(g_j+1)!\delta_{\alpha_j,g_j+1}\right),
\end{equation}
which can also be put in the form

\begin{equation}
\mathcal{C}(Q,N)=\sum_{i=0}^Q\binom{i+N-1}{i}
 \sum_{\vec{\alpha}/\sum_{j=1}^N\alpha_j=Q-i}\frac{1}{\alpha_1!\alpha_2!...\alpha_N!}
 \prod_{j=1}^N\left(\delta_{\alpha_j,0}-(g_j+1)!\delta_{\alpha_j,g_j+1}\right),
\end{equation}
which is the first main result of the present work.

\section{Second exact expression: grouping the supershells of the same degeneracy}\label{sec3}

Let us consider the case where $n_1$ orbitals have the same degeneracy $g_1$ and $n_2$ orbitals have the same degeneracy $g_2$, with $N=n_1+n_2$. For instance $(2p3p4p)^4$ and $(3d4d)^6$ correspond to $g_1$=6, $g_2$=10, $n_1$=4 and $n_2$=6, i.e. $N$=10. The generating function can be put in the form:

\begin{equation}
G(x,N)=\left(\frac{1-x^{g_1+1}}{1-x}\right)^{n_1}\left(\frac{1-x^{g_2+1}}{1-x}\right)^{n_2}.
\end{equation}

\noindent Using the Leibniz formula for the derivative of a product of two functions, we get

\begin{equation}
\mathcal{C}\left(Q,N\right)=\left.\frac{1}{Q!}\sum_{i=0}^Q\binom{Q}{i}\frac{\partial^i}{\partial x^i}\frac{1}{(1-x)^{n_1+n_2}}\right|_{x=0}
 \times\frac{\partial^{Q-i}}{\partial x^{Q-i}}\left[\left(1-x^{g_1+1}\right)^{n_1}
 \left.\left(1-x^{g_2+1}\right)^{n_2}\right]\right|_{x=0}.
\end{equation}

\noindent We still have

\begin{equation}
\left.\frac{\partial^i}{\partial x^i}\frac{1}{(1-x)^{n_1+n_2}}\right|_{x=0}=i!\binom{i+n_1+n_2-1}{i}
\end{equation}
and since
\begin{equation}
\left(1-x^{g_1+1}\right)^{n_1}=\sum_{i_1=0}^{n_1}(-1)^{i_1}\binom{n_1}{i_1}x^{i_1\left(g_1+1\right)},
\end{equation}
one can write
\begin{equation}\begin{split}
 &\left.\frac{\partial^{Q-i}}{\partial x^{Q-i}}\left[\left(1-x^{g_1+1}\right)^{n_1}
 \left(1-x^{g_2+1}\right)^{n_2}\right]\right|_{x=0}=\\
 &\sum_{i_1=0}^{n_1}\sum_{i_2=0}^{n_2}(-1)^{i_1+i_2}\binom{n_1}{i_1}\binom{n_2}{i_2}\\
 &\times\left[i_1\left(g_1+1\right)+i_2\left(g_2+1\right)-Q+i+1\right]
 \left[i_1\left(g_1+1\right)+i_2\left(g_2+1\right)-Q+i+2\right] \cdots\\
 &\times\left[i_1\left(g_1+1\right)+i_2\left(g_2+1\right)-2\right]
 \left[i_1\left(g_1+1\right)+i_2\left(g_2+1\right)-1\right]\\
 &\times\left[i_1\left(g_1+1\right)+i_2\left(g_2+1\right)\right]
 \left.\times x^{i_1\left(g_1+1\right)+i_2\left(g_2+1\right)-Q+i}\right|_{x=0}.
\end{split}\end{equation}

\noindent The only non-zero value on the right-hand side corresponds to $i=Q-i_1\left(g_1+1\right)-i_2\left(g_2+1\right)$ and we finally get

\begin{equation}
\mathcal{C}\left(Q,N\right)=\sum_{i_1=0}^{n_1}\sum_{i_2=0}^{n_2}(-1)^{i_1+i_2}\binom{n_1}{i_1}
 \binom{n_2}{i_2}\binom{n_1+n_2-1+Q-i_1\left(g_1+1\right)-i_2\left(g_2+1\right)}{n_1+n_2-1}.
\end{equation}

\noindent If we generalize and gather the $n_1$ subshells of degeneracy $g_1$, the $n_2$ subshells of degeneracy $g_2$, ..., the $n_s$ subshells of degeneracy $g_s$ (with therefore $n_1+n_2+\cdots n_s=N$), we obtain

\begin{equation}\begin{split}\label{eq:numberconf2}
\mathcal{C}\left(Q,N\right) =&\sum_{i_1=0}^{n_1}\sum_{i_2=0}^{n_2}\cdots\sum_{i_s=0}^{n_s}
 (-1)^{i_1+i_2+\cdots n_s}\binom{n_1}{i_1}\binom{n_2}{i_2}\cdots\binom{n_s}{i_s} \\
 &\times\binom{n_1+\cdots+n_s-1+Q-i_1\left(g_1+1\right)-i_2\left(g_2+1\right)-\cdots-i_s\left(g_s+1\right)}{n_1+\cdots+n_s-1},\\
\end{split}\end{equation}
which is the second main result of the present work.

\section{Recurrence relations on the number of subshells with same degeneracy}
\label{sec:rec_sameg}
The equation (\ref{eq:numberconf2}) is rather compact and adapted to numerical 
computation. However one may note that it contains terms of alternating signs. 
It is possible to derive an alternate formula containing only positive terms. 
Let us note $\mathscr{N}(N_1,\cdots N_t;g_1,\cdots g_t;Q)$ the 
number of configurations of $Q$ electrons distributed within $N_1$ distinct 
subshells of degeneracy $g_1$,\dots $N_t$ subshells of degeneracy $g_t$. 
For instance considering the non relativistic configurations constructed on 
the 1s 2s 2p 3s 3p 3d subshells , one has $t=3$, $N_1=3$, $g_1=2$, $N_2=2$, 
$g_2=6$, and $N_3=1$, $g_3=10$. It is clear that the evaluation of this 
number can be reduced to the evaluation of the number of the configurations of 
a given degeneracy $\mathscr{S}(g;N;Q)$ which is the number of configurations 
with $Q$ electrons distributed on $N$ subshells of the same degeneracy $g$. 
The numbers $\mathscr{N}$ and $\mathscr{S}$ are connected through the discrete 
convolution formula
\begin{equation}\label{eq:combinating_gs}
\mathscr{N}(N_1,\cdots N_t;g_1,\cdots g_t;Q)=
\sum_{p_1}\cdots\sum_{p_t}\delta_{p_1+\cdots+p_t,Q}
 \mathscr{S}(g_1;N_1;p_1)\cdots\mathscr{S}(g_t;N_t;p_t).
\end{equation}
In this section we will focus on the computation of the $\mathscr{S}(g;N;Q)$ numbers. 
Let us consider for instance the case $g=4$. To each configuration corresponds 
a 5-uple $(n_0,n_1,n_2,n_3,n_4)$ of numbers of subshells with population from 0 
to 4 respectively. Obviously two configurations with distinct 5-uples are 
different. Conversely, there are several distinct configurations for a given 
set $(n_0,n_1,n_2,n_3,n_4)$, that can be straightforwardly numbered. One has 
$\binom{N}{n_4}$ ways to choose the subshell(s) with 4 electrons, then 
$\binom{N-n_4}{n_3}$ ways to choose the remaining subshell(s) with 3 
electrons, etc. Therefore the total number of configurations writes
\begin{subequations}\begin{equation}
\left.\mathscr{S}(g;N;Q)\right|_{g=4}=
 \sum\binom{N}{n_4}\binom{N-n_4}{n_3}\binom{N-n_3-n_4}{n_2}
 \binom{N-n_2-n_3-n_4}{n_1}\binom{N-n_1-n_2-n_3-n_4}{n_0}
\end{equation}
where the summation is performed on all $(n_0,n_1,n_2,n_3,n_4)$ verifying
\begin{align}
N=&n_0+n_1+n_2+n_3+n_4\\
  Q=&n_1+2n_2+3n_3+4n_4.\end{align}
\end{subequations}
The product of binomial coefficients in the above sum simplifies, and one gets 
in the general case,
\begin{subequations}\begin{equation}\label{eq:Smultinom}
\mathscr{S}(g;N;Q)=\sum_{n_0,n_1\cdots n_g} \delta_{n_0+\cdots+n_g,N}
 \delta_{n_1+\cdots+gn_g,Q}\frac{N!}{n_0!n_1!\cdots n_g!}
\end{equation}
which, introducing the multinomial coefficient (\ref{eq:multinomial}), writes
\begin{equation}\label{eq:Smultinom2}
\mathscr{S}(g;N;Q)=\sum_{\stackrel{n_0,n_1\cdots n_g}{\mathscr{C}}}
 \binom{N}{n_0,n_1,\cdots n_g}
\end{equation}
where the multiple sum is constrained by the double condition $\mathscr{C}$
\begin{align}\label{eq:subshpop_constraint}
 N=&n_0+n_1+\cdots+n_g\\
 Q=&n_1+2n_2+\cdots+gn_g.
\end{align}\end{subequations}
This equation, in conjunction with (\ref{eq:combinating_gs}), provides a 
third expression for the total number of configurations. 
Let us now consider the generating function
\begin{subequations}\begin{align}
\mathscr{G}(g;z,X)=&\sum_{n_0=0}^\infty\frac{z^{n_0}}{n_0!}
 \sum_{n_1=0}^\infty\frac{z^{n_1}X^{n_1}}{n_1!}
 \sum_{n_2=0}^\infty\frac{z^{n_2}X^{2n_2}}{n_2!}
 \cdots\sum_{n_g=0}^\infty\frac{z^{n_g}X^{gn_g}}{n_g!}\\
 =& \exp(z+zX+zX^2\cdots+zX^g)\label{eq:fgenXzexpanded}\\
 =& \exp\left(z\frac{1-X^{g+1}}{1-X}\right).\label{eq:fgenXzcompact}
\end{align}\end{subequations}
Comparing the above expansion with the value (\ref{eq:Smultinom}) one 
checks that
\begin{equation}\label{eq:GenfunczX}
 \mathscr{G}(g;z,X)=\sum_{Q=0}^\infty
 \sum_{N=0}^\infty\mathscr{S}(g;N;Q) \frac{z^N}{N!} X^Q.
\end{equation}
Therefore one may express the number of configurations as the partial derivative
\begin{equation}
 \mathscr{S}(g;N;Q) = \frac{1}{Q!}  \left.\frac{\partial^{N+Q}}{\partial 
  z^N\partial X^Q} \mathscr{G}(g;z,X)\right|_{z=0,X=0}.
\end{equation}
The above expansion allows us to derive various properties. Using the form 
(\ref{eq:fgenXzcompact}) one easily verifies that
\begin{equation}
 \mathscr{G}(g;z,X)=\mathscr{G}(g;zX^g,1/X)
\end{equation}
which implies
\begin{equation}\label{eq:symmetryS}
 \mathscr{S}(g;N;Q)=\mathscr{S}(g;N;gN-Q).
\end{equation}
Recursion relations can be obtained by deriving the generating function 
(\ref{eq:fgenXzcompact}) with respect to $z$ or $X$. Writing the ratio 
$(1-X^{g+1})/(1-X)$ (resp. its derivative) as the polynomial $1+X+\cdots+X^g$ 
(resp. $1+2X+\cdots+gX^{g-1}$), one gets two identities. First, using 
derivation versus $z$ and identifying terms in $z^N X^{Q}$ one has
\begin{equation}\label{eq:recurS1}
\mathscr{S}(g;N+1;Q)=\sum_{j=0}^{\min(g,Q)}\mathscr{S}(g;N;Q-j).
\end{equation}
Then, using derivation versus $X$, assuming $Q>0$, one obtains
\begin{equation}\label{eq:recurS2}
\mathscr{S}(g;N+1;Q)=\frac{N+1}{Q}
 \sum_{j=1}^{\min(g,Q)}j\mathscr{S}(g;N;Q-j).
\end{equation}
In a similar way, dealing with $(1-X^{g+1})/(1-X)$ or its derivative as a 
rational fraction one first gets by deriving with respect to $z$ 
\begin{subequations}\begin{align}
 \frac{1-X^{g+1}}{1-X}\exp\left(z\frac{1-X^{g+1}}{1-X}\right)
 =&\frac{1-X^{g+1}}{1-X}\sum_{N,Q}\mathscr{S}(g;N;Q)\frac{z^N}{N!} X^Q\\
 =&\sum_{N,Q}\mathscr{S}(g;N;Q)\frac{z^{N-1}}{(N-1)!} X^Q
\end{align}
and after multiplying the right-hand sides of these subequations by $(1-X)$ 
and identifying the factor of $z^N X^Q$, one has
\begin{equation}\label{eq:recurS3}
 \mathscr{S}(g;N+1;Q)-\mathscr{S}(g;N+1;Q-1)=
  \mathscr{S}(g;N;Q)-\mathscr{S}(g;N;Q-g-1).
\end{equation}\end{subequations}
Then after deriving the generating function $\mathscr{G}$ with respect to 
$X$ and multiplying both sides by $(1-X)^2$, 
\begin{subequations}\begin{equation}
 z\left[1-(g+1)X^g+gX^{g+1}\right]\sum_{N Q}\mathscr{S}(g;N;Q)
  \frac{z^N}{N!} X^Q
  =(1-X)^2\sum_{N Q} Q\mathscr{S}(g;N;Q)\frac{z^N}{N!} X^{Q-1}
\end{equation}
and term-by-term identification leads to the recurrence relation 
\begin{multline}\label{eq:recurS4}
 (Q+1)\mathscr{S}(g;N+1;Q+1)-2Q\mathscr{S}(g;N+1;Q)+
      (Q-1)\mathscr{S}(g;N+1;Q-1)\\=
  (N+1)\Big(\mathscr{S}(g;N;Q)-(g+1)\mathscr{S}(g;N;Q-g)
  +g\mathscr{S}(g;N;Q-g-1)\Big).
\end{multline}\end{subequations}
The first recurrence (\ref{eq:recurS1}) has been mentioned previously 
(\ref{eq:recurnbconfig}). If the $\mathscr{S}(g;N;Q)$ numbers are written 
in a Pascal-like triangle where lines are indexed by $N$ and columns by 
$Q$, this equation implies that any number in the array is equal to the sum 
of the numbers located on the row above at the $g+1$ positions ending at 
the current column --- ignoring elements with negative column indices. 
In the special case $g=1$ this rule reverts to the usual triangle rule so 
that
\begin{equation}\label{eq:Pascal}
\mathscr{S}(1;N;Q)=\binom{N}{Q}.
\end{equation}
Of course this relation could also have been obtained by a direct argument. 
Noting that the generating function (\ref{eq:GenfunczX}) verifies
\begin{equation}
 \mathscr{G}(g;z,X)=\mathscr{G}(g-1;z,X)\exp(zX^g)
\end{equation}
one obtains an additional recurrence relation on the degeneracy $g$. This 
equation may be written, with the above definitions
\begin{equation}
 \sum_{N Q}\mathscr{S}(g;N;Q)\frac{z^N}{N!}X^Q = 
  \mathscr{S}(g-1;N;Q)\sum_j\frac{z^j X^{jg}}{j!}
\end{equation}
and identifying the terms in $z^N X^Q$ on both sides one gets 
\begin{equation}\label{eq:recurr_g}
 \mathscr{S}(g;N;Q) = 
 \sum_j \binom{N}{j}\mathscr{S}(g-1;N-j;Q-jg)
\end{equation}
the minimum index $j$ being $\max(0,Q-(g-1)N)$ so that one has $Q-jg\le
(g-1)(N-j)$, and the maximum index $j$ being $\min(N,\lfloor Q/g\rfloor)$.
With the initial value (\ref{eq:Pascal}), this relation may be used to get all 
$\mathscr{S}(g;N;Q)$. Because of the symmetry property (\ref{eq:symmetryS}), 
for a given number of subshells $N$ the evaluation needs only to be done for 
$0\le Q\le p_\text{max}=\lfloor (gN+1)/2\rfloor$. For low $Q$ values, the 
sum (\ref{eq:recurr_g}) contains very few terms since one must have $Q-jg\ge0$. 
For $Q=p_\text{max}$, the maximum index $j$ is only $\lfloor(N+1)/2\rfloor$.

Up to our knowledge, the recurrence relations (\ref{eq:recurS2},
\ref{eq:recurS3},\ref{eq:recurS4},\ref{eq:recurr_g}) have not been published 
previously. Using a batch of test values (mostly in the $g=6$ case) we have 
checked that the various recurrences obtained here are numerically correct. 
Moreover, at variance with the relations derived in the previous sections, 
the sums in the right-hand side of (\ref{eq:recurS1},\ref{eq:recurS2}) 
involve only positive terms and therefore cannot give rise to a loss of 
accuracy or instability after repeated use of the recurrence.

\section{Analysis of the distribution of populations among $N$ distinct 
subshells with the same degeneracy using moments calculation}
\label{sec:moments}
The formulas given in the preceding sections, and mostly those involving 
recurrence relations, provide a very fast method to get a large set of 
$\mathscr{S}(g;N;Q)$ values. As mentioned before, if $g=1$ the 
distribution of $\mathscr{S}$ as a function of $Q$ is binomial. A very 
efficient characterization of such distributions lies in the analysis of 
moments defined, for a given degeneracy $g$ and subshell number $N$, as
\begin{equation}\label{eq:moment}
\mathscr{M}(g;N;k) = \sum_{Q=0}^{gN} Q^k\mathscr{S}(g;N;Q).
\end{equation}
The moment analysis is, in particular, crucial in the study of \emph{unresolved 
transition arrays} as proven by Bauche \textit{et al.} 
\cite{BAUCHE15}. It allows to give a simple and often accurate 
description of such arrays through the definition of a small number of such 
moments.

We have been able to derive analytically or numerically the corresponding 
formulas for the moments. Indeed, it has been mentioned in several works 
\cite{Pain2009a,Pain2010,Na2016} that, in some cases, the knowledge of the 
moments up to the second (variance) is far from sufficient to describe 
distributions significantly different from the normal distribution. This is 
why a certain effort is devoted here to moments up to a quite large order.

First one easily finds that
\begin{equation}\label{eq:M0}\mathscr{M}(g;N;0) = (g+1)^N\end{equation}
since this is the total number of configurations with any number of electrons 
distributed over $N$ subshells of degeneracy $g$. As mentioned in Eq. 
(\ref{eq:symmetryS}) the $\mathscr{S}(g;N;Q)$ distribution is symmetric 
with respect to its median value $gN/2$, and this provides immediately the 
next moment 
\begin{equation}\mathscr{M}(g;N;1) = \frac12gN(g+1)^N.\end{equation}

\noindent The generating function (\ref{eq:GenfunczX}) also allows us to derive expressions 
for moments at any order in a closed form. Explicitly, one has for the $k$-th 
order derivative with respect to $X$
\begin{equation}
\frac{\partial^k\mathscr{G}(g;z;X)}{\partial X^k}= \sum_{Q=0}^\infty
 \sum_{n=0}^\infty (Q)_k\mathscr{S}(g;n;Q) \frac{z^n}{n!} X^{Q-k}.
\end{equation}
where, for integer $n$,
\begin{equation}(A)_n=A(A-1)...(A-n+1)\end{equation}
is the so-called descending factorial. 
Evaluating this quantity for $X=1$ provides the successive moments of the 
$\mathscr{S}$ distribution. Indeed, one easily checks using the analytical form 
(\ref{eq:fgenXzexpanded})
\begin{subequations}\begin{equation}
\left.\frac{\partial^k}{\partial X^k}\exp\left(z(1+X+\cdots+X^g)\right)\right|_{X=1}
= \sum_{z=0}^\infty \frac{z^N}{N!}\overline{\mathscr{M}}(g;N;k)
\end{equation}
with \begin{equation}
\overline{\mathscr{M}}(g;N;k)=\sum_{Q=0}^{gN}(Q)_k\mathscr{S}(g;N;Q).
\end{equation}
Therefore the modified moments 
$\overline{\mathscr{M}}$ appear as the $(N+k)$-th partial derivative
\begin{align}\label{eq:Mk_derN+k}
\overline{\mathscr{M}}(g;N;k)=& \left.\frac{\partial^{N+k}}{\partial X^k\partial z^N}
 \exp\left(z(1+X+\cdots+X^g)\right)\right|_{X=1,z=0}\\
 =& \left.\frac{\partial^{k}}{\partial X^k}
 \left[(1+X+\cdots+X^g)^N\exp\Big(z(1+X+\cdots+X^g)\Big)\right]
 \right|_{X=1,z=0}.\label{eq:MderNderk}
\end{align}
\end{subequations}
For instance, in the case $k=0$, one gets immediately $(g+1)^N$ as mentioned above 
(\ref{eq:M0}). The various moments (\ref{eq:moment}) can be easily related to the 
sums obtained above (\ref{eq:MderNderk}) since one has 
\begin{equation}\label{eq:Stirling_fallfact}
x^n=\sum_{j=0}^n\left\{\begin{matrix}n\\j\end{matrix}\right\}(x)_j
\end{equation}
the coefficients on the right-hand side being the Stirling numbers of the second 
kind \cite{Comtet1974}. 
These numbers can be easily generated from the recurrence \cite{Abramowitz1972}
\begin{equation}
\left\{\begin{matrix}n+1\\m\end{matrix}\right\}
=m\left\{\begin{matrix}n\\m\end{matrix}\right\}+
\left\{\begin{matrix}n\\m-1\end{matrix}\right\}\text{ with, by convention, }
\left\{\begin{matrix}0\\0\end{matrix}\right\}=1, 
\left\{\begin{matrix}n\\0\end{matrix}\right\}=0\text{ if }n>0.
\end{equation}
Furthermore, the Arbogast-Faà di Bruno's formula allows us to write 
\cite{Abramowitz1972,Comtet1974}
\begin{equation}
\label{eq:derk_SN}
\frac{\partial^k S(X)^N}{\partial X^k} =
 \sum_{n_1,n_2,\cdots,n_k} \delta_{k,n_1+2n_2\cdots+kn_k}
 \mathscr{P}(k;n_1,n_2\cdots,n_k) 
 S^{(1)}(X)^{n_1}S^{(2)}(X)^{n_2}\cdots S^{(k)}(X)^{n_k} (N)_d S(X)^{N-d}
\end{equation}
where 
\begin{equation}\label{eq:sumnj}d=n_1+n_2+\cdots+n_k\end{equation} 
and where 
$\mathscr{P}(k;n_1,n_2\cdots,n_k)$ is the number of partitions of $k$ distinct 
objects with $n_1$ groups containing 1 element, $n_2$ groups containing 2 
elements,\dots $n_k$ groups containing $k$ elements. The number $\mathscr{P}$ 
is given by Eq. (\ref{eq:partition_number}) of Appendix \ref{sec:numparts}. 
In order to close the computation, one needs to substitute $1+X+\cdots +X^g$ 
to $S(X)$ in the derivative formula (\ref{eq:derk_SN}) and therefore to 
compute the partial derivative
\begin{equation}
\mathscr{T}_j=\left.\frac{\partial^j}{\partial X^j}(1+X+\cdots X^g)\right|_{X=1}.
\end{equation}
This can be easily performed by explicitly deriving the first values 
\begin{equation}\mathscr{T}_0=g+1, \mathscr{T}_1=\frac12g(g+1), 
\mathscr{T}_2=\frac13(g-1)g(g+1),\end{equation}
from which one infers the general form
\begin{equation}\label{eq:derelem}
\mathscr{T}_r=\frac{1}{r+1}\frac{(g+1)!}{(g-r)!}
 =r!\binom{g+1}{r+1}.\end{equation}
The proof of the above alleged expression can be established by a simple 
recurrence on the index $g$. The average over the distribution $\mathscr{S}(g;N;Q)$ of any function of $Q$ 
$\mathscr{X}(Q)$ is defined as 
\begin{equation}
\left<\mathscr{X}(Q)\right>=\sum_Q\mathscr{X}(Q)\mathscr{S}(g;N;Q)/
 \sum_Q\mathscr{S}(g;N;Q)=\sum_Q\mathscr{X}(Q)\mathscr{S}(g;N;Q)/(g+1)^N.
 \label{eq:average_X_Q}
\end{equation}
Collecting formulas (\ref{eq:Mk_derN+k},\ref{eq:derk_SN},\ref{eq:derelem},
\ref{eq:average_X_Q},\ref{eq:partition_number}), and noting that the factor 
$S(X)^{N-d}$ in Eq. (\ref{eq:derk_SN}) may be written as
\begin{equation}
S(X)^{N-d} = (g+1)^{N-n_1-n_2\cdots -n_k},
\end{equation}
one gets finally the average value of the descending factorials $(Q)_k$, 
\begin{equation}\label{eq:aver_fall_fact}
\left<(Q)_k\right> = \frac{\overline{\mathscr{M}}(g;N;k)}{(g+1)^N}
 \sum_{n_1\cdots n_k} 
 \frac{ \delta_{j,n_1+2n_2+\cdots kn_k} k!}{\prod_{q=1}^k n_q!(q!)^{n_q}}
 (N)_{n_1+n_2\cdots+n_k}
 \prod_{r=1}^k\left[\frac{1}{r+1}\frac{g!}{(g-r)!}\right]^{n_r}
\end{equation}
and the normalized moments, using the sum (\ref{eq:Stirling_fallfact}),
\begin{equation}\label{eq:momentS_nc}
\mathscr{M}(g;N;k)/(g+1)^N = 
 \sum_{j=0}^k\left\{\begin{matrix}k\\j\end{matrix}\right\}
 \sum_{n_1\cdots n_j} 
 \frac{ \delta_{j,n_1+2n_2+\cdots jn_j} j!}{\prod_{q=1}^j n_q!(q!)^{n_q}}
 (N)_{n_1+n_2\cdots+n_j}
 \prod_{r=1}^j\left[\frac{1}{r+1}\frac{g!}{(g-r)!}\right]^{n_r}.
\end{equation}
Using the second form for $\mathscr{T}$ as written in Eq.~(\ref{eq:derelem}) 
one may also write the somewhat simpler result
\begin{equation}\label{eq:aver_fall_fact2}
\left<(Q)_k\right>/(g+1)^N = 
 \sum_{n_1\cdots n_k} \delta_{k,n_1+2n_2+\cdots kn_k}  
 \frac{k!}{\prod_{q=1}^k n_q!}\:
 \frac{(N)_{d}}{(g+1)^d} \prod_{r=1}^k\binom{g+1}{r+1}^{n_r}
\end{equation}
with $d$ is the sum of the $n_j$ indices (\ref{eq:sumnj}).

The moments with $k\le 8$ have been explicitly obtained and are listed in 
table \ref{tab:momt_distrS}. The formulas have been obtained using Mathematica 
software, though the lowest moments may be easily derived by using the explicit 
form (\ref{eq:momentS_nc}). We have checked that, in spite of the multiple 
nested loops on indices $n_j$ in the expression (\ref{eq:aver_fall_fact}), the 
analytical expressions for moments up to $k=10$ can be obtained at a very low 
computational cost. Indeed, considering for instance the 4-th order moment, 
the nested loop on $n_j$ indices only contains four terms, namely $(n_1=4), 
(n_1=2,n_2=1), (n_1=1,n_3=1), (n_4=1)$, where all the unmentioned $n_j$ 
are 0. From the above expression one may also notice that 
each of these normalized moments is given by a polynomial form
\begin{equation}\label{eq:coefpol_moment}
 \mathscr{M}(g;N;k)/(g+1)^N = \sum_{p=1}^k\sum_{q=1}^k c_{pq}(g;N;k)g^pN^q.
\end{equation}
To get moments for large $k$ values, it may be easier to use such formula instead of 
(\ref{eq:momentS_nc}). One first computes numerically a series of moments for various 
$g$ and $N$ values using the previously mentioned recurrence relation, and one then 
solves the linear system (\ref{eq:coefpol_moment}) to obtain the $c_{pq}$.

\begin{table}[htbp]
\centering
\renewcommand{\arraystretch}{2.5}
\begin{tabular}{c@{\qquad}l}
\hline\hline
$k$ & Non-centred moment\\[0.5ex]
\hline
 2 & $\displaystyle\frac14g^2 N^2 + \frac{1}{12}g(g+2)N$ \\
 3 & $\displaystyle\frac18g^3 N^3 + \frac{1}{8}g^2(g+2)N^2$ \\
 4 & $\displaystyle\frac{1}{16}g^4 N^4 + \frac{1}{8}g^3(g+2) N^3
  + \frac{1}{48}g^2(g+2)^2 N^2 - \frac{1}{120}g(g+2) (g^2+2g+2) N$ \\
 5 & $\displaystyle\frac{1}{32}g^5 N^5 + \frac{5}{48} g^4(g+2)N^4
  + \frac{5}{96}g^3(g+2)^2 N^3 - \frac{1}{48}g^2(g+2)(g^2+2g+2) N^2$ \\
 6 & $\begin{aligned}[t]&\frac{1}{64}g^6 N^6 + \frac{5}{64}g^5(g+2)N^5 
  + \frac{5}{64}g^4(g+2)^2 N^4 - \frac{1}{576}g^3(g+2)(13g^2+16g+16)N^3 \\
  &\quad - \frac{1}{96}g^2(g+2)^2(g^2+2g+2)N^2
    + \frac{1}{252}g(g+2)(g^2+g+1)(g^2+3g+3)N\end{aligned}$ \\
 7 & $\begin{aligned}[t]&\frac{1}{128}g^7 N^7 + \frac{7}{128} g^6(g+2)N^6 
 + \frac{35}{384}g^5(g+2)^2N^5 - \frac{7}{1152}g^4(g+2)(g^2-8g-8)N^4 \\
 &\quad - \frac{7}{192}g^3(g+2)^2(g^2+2g+2)N^3 
 + \frac{1}{72}g^2(g+2)(g^2+g+1)(g^2+3g+3)N^2\end{aligned}$\\
 8 & $\begin{aligned}[t]&\frac{1}{256}g^8 N^8 + \frac{7}{192}g^7(g+2)N^7 
 + \frac{35}{384}g^6(g+2)^2 N^6 + \frac{7}{288}g^5(g+2)(g^2+7g+7) N^5 \\
 &\quad - \frac{7}{6912}g^4(g+2)^2(67g^2+124g+124) N^4
 + \frac{1}{576}g^3(g+2)(9g^4+22g^3+14g^2-16g - 8) N^3\\
 &\quad + \frac{1}{8640}g^2(g+2)^2(101g^4+404g^3+728g^2+648g+324) N^2\\
 &\quad - \frac{1}{240}g(g+2)(g^2+2g+2)(g^4+4g^3+6g^2+4g+2) N\strutl
 \end{aligned}$ \\
\hline\hline
\end{tabular}
\caption{Normalized non-centered moments $\mathscr{M}(g;N;k)/(g+1)^N$ of 
the distribution $\mathscr{S}(g;N;Q)$.\label{tab:momt_distrS}}
\end{table}

\noindent The \emph{normalized centred moments} are defined as
\begin{equation}
\mathfrak{M}_c(g;N;k)=\sum_{Q=0}^{gN}\left(Q-gN/2\right)^k\mathscr{S}(g;N;Q)/(g+1)^N.
\end{equation}
Because of the symmetry property (\ref{eq:symmetryS}), these moments cancel if $k$ 
is odd. Using the above equation and the general expression (\ref{eq:momentS_nc}) one
obtains moments with a somewhat simpler form than the non-centred moments. 
Namely one has 
\begin{subequations}\label{eq:cmts2-10}\begin{align}
\mathfrak{M}_c(g;N;2) &= \frac{1}{12}g(g+2) N 
  =\frac{D_2}{12} N \label{eq:variance}\\
\mathfrak{M}_c(g;N;4) &= \frac{1}{48}g^2(g+2)^2 N^2 - \frac{1}{120}g(g+2)(g^2+2g+2) N
  = \frac{D_2^2}{48} N^2 - \frac{D_4}{120} N
\label{eq:kurtosis}\\
\mathfrak{M}_c(g;N;6) &= \frac{5}{576}g^3(g+2)^3 N^3 - \frac{1}{96}g^2(g+2)^2(g^2+2g+2) N^2 
 + \frac{1}{252}g(g+2)(g^2+g+1)(g^2+3g+3) N \nonumber\\
  &= \frac{5D_2^3}{576}N^3 -\frac{D_2D_4}{96}N^2 + \frac{D_6}{252}N
  \label{eq:superkurtosis}\\
\mathfrak{M}_c(g;N;8) &= \frac{35}{6912}g^4(g+2)^4 N^4 
 - \frac{7}{576}g^3(g+2)^3(g^2+2g+2) N^3 \nonumber\\
 &\quad +\frac{1}{8640}g^2(g+2)^2(101g^4+404g^3+728g^2+648g+324) N^2 \nonumber\\
 &\quad -\frac{1}{240}g(g+2)(g^2+2g+2)(g^4+4g^3+6g^2+4g+2) N\nonumber\\
 &= \frac{35D_2^4}{6912}N^4-\frac{7D_2^2D_4}{576}N^3
  +\frac{D_2D_6}{108}N^2+\frac{7D_4^2}{2880}N^2-\frac{D_8}{240}N.
\end{align}
where we have defined, for the sake of simplification,
\begin{equation}D_k=(g+1)^k-1.\end{equation}
\end{subequations}

\noindent The first centred moment of this list is the variance
\begin{equation}\sigma^2=\frac{1}{12}g(g+2) N\label{eq:varS}
\end{equation}
and the second one is related to the excess kurtosis, given by \cite{Stuart1994}
\begin{equation}
 \kappa_4/\sigma^4 = \mathfrak{M}_c(g;N;4)/\sigma^4 -3 
  = -\frac{6(g^2+2g+2)}{5g(g+2)N}
\end{equation}
which would be zero for a normal distribution. As one will verify below, the 
excess kurtosis can be significantly different from 0, especially for large 
$g$ and moderate $N$. This negative value means that such distributions, 
named platykurtic, are flatter than the normal distribution. 
Conversely, for a given $g$, one has $\lim_{N\to\infty}\kappa_4/\sigma^4=0$.

\section{Cumulant analysis}\label{sec:cumulants}
The previous considerations are useful to characterize the $\mathscr{S}$ 
population distribution, e.g., by comparing it to a normal distribution. 
They can be used to compute Gram-Charlier approximations (by truncating 
this series at various orders). However they suffer from two limitations. 
The first one is that the expressions for the moments increase in complexity 
with the order $k$. The second one is that they do not apply when several 
subshells with different degeneracies $g$ are present in the supershell. 

To circumvent these limitations, one must resort to the \emph{cumulant} 
formalism. The global distribution is given by the discrete convolution 
formula (\ref{eq:combinating_gs}). While the normalized centred moments 
cannot in the general case be expressed as the sum of the $(g_j,n_j)$ moments 
(\ref{eq:cmts2-10}), the \emph{additivity holds for the cumulants}. 

The \emph{generating function} for the cumulants is defined as 
\cite{Stuart1994}
\begin{subequations}\begin{align}\label{eq:somcumul}
\mathscr{K}(t) &= \sum_{n=1}^\infty\kappa_n\frac{t^k}{n!}\\
 &= \log\left(\left< \exp(tQ) \right>\right).\label{eq:defcumul}
\end{align}\end{subequations}
Considering first the case of $N$ distinct subshells with the same degeneracy 
$g$, the above average value $\left<\exp(tQ)\right>$ can be easily computed. 
Using the well-known property arising from the convolution relation 
(\ref{eq:recurnbconfig})
\begin{equation}
\mathscr{S}(g;N_1+N_2;Q)=\sum_{j}\mathscr{S}(g;N_1;j)\mathscr{S}(g;N_2;Q-j)
\end{equation}
one has for the Laplace-transformed expression for any natural integers 
$N_1$, $N_2$
\begin{subequations}\begin{align}
\sum_Q\mathscr{S}(g;N_1+N_2;Q)e^{Qt}&= 
 \sum_{Q,j}\mathscr{S}(g;N_1;j)e^{jt}\mathscr{S}(g;N_2;Q-j)e^{(Q-j)t}\\
 &= \left(\sum_j\mathscr{S}(g;N_1;j)e^{jt}\right)
     \left(\sum_k\mathscr{S}(g;N_2;k)e^{kt}\right)
\end{align}\end{subequations}
and by repeated application of the convolution formula
\begin{equation}
\sum_Q\mathscr{S}(g;N;Q)e^{Qt} =
 \left(\sum_j\mathscr{S}(g;1;j)e^{jt}\right)^N.
\end{equation}
The sum raised to the $N$-th power is evaluated straightforwardly. Using the 
$N=1$ value
\begin{equation}\mathscr{S}(g;1;Q)=\theta(g-Q)\label{eq:SgN1Q}\end{equation}
which comes directly from the definition of $\mathscr{S}$, one gets
\begin{equation}
\sum_j\mathscr{S}(g;1;j)e^{jt}=1+e^t+\cdots+e^{gt}=\frac{e^{ht}-1}{e^t-1}
 =\exp\left(\frac{gt}{2}\right)\frac{\sinh(ht/2)}{\sinh(t/2)}
\end{equation}
where $h=g+1$. Using the normalization relation (\ref{eq:M0}), one obtains 
the average value for the case with $N$ distinct subshells with the same 
degeneracy 
\begin{equation}
\left< e^{Qt} \right> = 
 \sum_Q\mathscr{S}(g;N;Q)e^{Qt} \left/ \sum_Q\mathscr{S}(g;N;Q)\right. 
 =e^{Ngt/2}\left( \frac{\sinh(ht/2)}{h\sinh(t/2)}\right)^N
\end{equation}
and considering the centred variable $Q-\left<Q\right>$ one has, since 
$\left<Q\right>=gN/2$,
\begin{equation}
\left< e^{(Q-\left<Q\right>)t} \right> = 
 \left( \frac{\sinh(ht/2)}{h\sinh(t/2)} \right)^N.
\end{equation}
Let us note that the above relations are formally equivalent to the ones 
providing the partition function of a quantum magnetic momentum interacting 
with a magnetic field in the theory of paramagnetism. In order to get the 
cumulants one must according to the definition (\ref{eq:defcumul}), compute 
the $k$-th derivative of the generating function 
\begin{equation}\label{eq:genK}
 \mathscr{K}(t) = N\log \left( \frac{\sinh(ht/2)}{h\sinh(t/2)} \right).
\end{equation}
These derivatives may be obtained by various methods. Let us now consider the 
Taylor series
\begin{subequations}\begin{align}
K(t) &=N\sum_{k=1}^\infty\frac{B_{2k}}{2k}\frac{h^{2k}-1}{(2k)!}t^{2k}\\
 &=N\left(\mathscr{G}(ht)-\mathscr{G}(t)\right)\label{eq:genKs}
\end{align}
where $B_j$ are the Bernoulli numbers and where
\begin{equation}
 \mathscr{G}(X)=\sum_{k=1}^\infty\frac{B_{2k}}{2k}\frac{X^{2k}}{(2k)!}.
\end{equation}
One gets, using a well known property of the Bernoulli numbers,
\begin{align}
\mathscr{G}(X)
 = \int_0^X\!\!\frac{du}{u}\sum_{k=1}^\infty B_{2k}\frac{u^{2k}}{(2k)!}
 = \int_0^X\!\!\frac{du}{u}
     \left(\frac12\coth\left(\frac{u}{2}\right)-\frac1u\right)
 = \log\left(\frac{2}{X}\sinh\left(\frac{X}{2}\right)\right).\label{eq:GX}
\end{align}\end{subequations}
Inserting formula (\ref{eq:GX}) in the above expression (\ref{eq:genKs}) for 
the cumulant generating function, and comparing the analytical expressions 
(\ref{eq:genK},\ref{eq:genKs}), one readily obtains
\begin{equation}\mathscr{K}(t)=K(t).\end{equation}
From the expansion (\ref{eq:somcumul}) one obtains directly the even-order 
cumulants 
\begin{equation}\label{eq:cumulantB}
 \kappa_{2k} = N\frac{B_{2k}}{2k}\left((g+1)^{2k}-1\right)
\end{equation}
in the case of a unique $g$ value. 
Because of the definition (\ref{eq:defcumul}), when subshells of various $g$ 
are involved, the average value $\left< e^{Qt} \right>$ is simply the product 
of the average on each subshell, the global $\mathscr{K}(t)$ is the sum of 
the individual generating functions, and the $2k$-th derivative provides the 
cumulant
\begin{equation}\label{eq:cumulant_gen}
 \kappa_{2k} = \frac{B_{2k}}{2k}\sum_j N_j\left((g_j+1)^{2k}-1\right)
\end{equation}
for the most general supershell.

Assuming $\mu_j$ are \emph{centred} moments, then $\kappa_1$ cancels, and the 
general relation giving moments as function of cumulants is \cite{Stuart1994}
\begin{equation}\label{eq:rel_momt-cumult}
\mu_n = \sum_{\substack{a_2\cdots,a_n\\2a_2\cdots+na_n=n}}
\mathscr{P}(n;a_2\cdots,a_n)\kappa_2^{a_2}\cdots\kappa_n^{a_n}
\end{equation}
where the coefficient $\mathscr{P}$ is defined in Appendix \ref{sec:numparts}. 
Since in the present case, all odd-order moments (or cumulants) cancel, one 
may limit the index sets to even-order sets $a_2, a_4\cdots a_{2k}$ with 
$n=2k$. As an example, defining 
\begin{equation}\label{eq:def_Ck_h}
 C_k=\sum_{j=1}^t(h_j^k-1)N_j\text{\quad with }h_j=g_j+1,
\end{equation}
one gets new expressions for the first centred moments
\begin{subequations}\begin{align}\label{eq:moments_Ck}
\mu_2 &= \frac{C_2}{12}\\
\mu_4 &= \frac{C_2^2}{48} -\frac{C_4}{120}\\
\mu_6 &= \frac{5C_2^3}{576} -\frac{C_2C_4}{96} +\frac{C_6}{252}\\
\mu_8 &= \frac{35 C_2^4}{6912} -\frac{7 C_4 C_2^2}{576} +\frac{C_6 C_2}{108}
 +\frac{7 C_4^2}{2880} -\frac{C_8}{240}\\
\mu_{10} &= \frac{35 C_2^5}{9216} -\frac{35 C_4 C_2^3}{2304} + 
 \frac{5 C_6 C_2^2}{288} +\frac{7 C_4^2 C_2}{768} -\frac{C_8 C_2}{64} 
 -\frac{C_4 C_6}{144} +\frac{C_{10}}{132}\\
\mu_{12} &= \frac{385 C_2^6}{110592} -\frac{385 C_4 C_2^4}{18432}
 +\frac{55 C_6C_2^3}{1728} +\frac{77 C_4^2 C_2^2}{3072} 
 -\frac{11 C_8 C_2^2}{256} -\frac{11 C_4 C_6 C_2}{288} \nonumber\\
 &\quad+\frac{C_{10} C_2}{24} -\frac{77 C_4^3}{23040} 
 +\frac{11 C_6^2}{1512} +\frac{11 C_4 C_8}{640} -\frac{691 C_{12}}{32760}
\end{align}\end{subequations}
which are more general than the previous ones (\ref{eq:cmts2-10}) since they 
apply to the case where several distinct $g_j$ are present.

\section{Analysis of population distribution with a Gram-Charlier expansion}
\label{sec:Gram-Charlier}
According to statistical treaties, any distribution such as 
(\ref{eq:combinating_gs}) may be approximated by a Gram-Charlier expansion, 
which is defined as (see Sec. 6.17 in Ref.\cite{Stuart1994})
\begin{equation}\label{eq:Gram-Charlier}
F_\text{GC}(Q) = \frac{G}{(2\pi)^{1/2}\sigma}
 \exp\left[-\frac{(Q-\left<Q\right>)^2}{2\sigma^2}\right] \left[1+\sum_{k\ge1}
 c_k He_k\left(\frac{Q-\left<Q\right>}{\sigma}\right)\right]
\end{equation}
where the $He_n$ is the Chebyshev-Hermite polynomial \cite{Stuart1994} 
\begin{equation}\label{eq:Hermite_stat}
He_k(X) = k!\sum_{m=0}^{\lfloor k/2\rfloor}\frac{(-1)^m X^{k-2m}}{2^m m! (k-2m)!}
\end{equation}
and $\lfloor x\rfloor$ is the integer part of $x$. 
The Gram-Charlier coefficients $c_k$ are related to the centred moments 
$\mu_k$ through the relation
\begin{equation}\label{eq:coeff_Gram-Charlier}
c_k=\sum_{j=0}^{\lfloor k/2\rfloor}\frac{(-1)^j
 \mu_{k-2j}/\sigma^{k-2j}}{2^j j!(k-2j)!}
\end{equation}
and from this definition the coefficients $c_1$ and $c_2$ cancel. For a 
symmetric distribution as the one considered here, all the odd-order terms 
$c_k$ cancel too. In the present case, the coefficient $G$ in 
Eq.(\ref{eq:Gram-Charlier}) is given by the normalization condition
\begin{equation}
G = \int_{-\infty}^{\infty}\!\!dQ\; F_\text{GC}(Q) =
 \prod_j\sum_{Q=0}^{N_jg_j}\mathscr{S}(g_j;N_j;Q) = \prod_j(g_j+1)^{N_j},
\end{equation}
the average value is $\left<Q\right>=\sum_j(g_jN_j)/2$ and the variance 
is $\sigma^2=\frac1{12}\sum_jg_j(g_j+2)N_j$.
As shown by Eq. (\ref{eq:cGC_vs_cumulants}) of Appendix 
\ref{sec:app_GC_vs_cumul}, one may also express the 
Gram-Charlier coefficients as a function of the cumulants. 

\subsection{Single-degeneracy case}
We first consider here the case where only one degeneracy $g$ is present.
In Eq. (\ref{eq:Gram-Charlier}), one chooses $\left<Q\right>=gN/2$ and 
$\sigma$ given by (\ref{eq:varS}). 
Using the general relation between $c_k$ coefficients and cumulants 
(\ref{eq:cGC_vs_cumulants}) and the cumulant value (\ref{eq:cumulantB}) one 
gets 
\begin{subequations}\label{eq:CGC4-10}\begin{align}
c_4 &=-\frac{h^2+1}{20(h^2-1)N}\label{eq:cGC4}\\
c_6 &=\frac{h^6-1}{105(h^2-1)^3 N^2}\\
c_8 &=-\frac{(h^2+1)\left[12(h^4+1)-7(h^4-1)N\right]}{5600 (h^2-1)^3 N^3}\\
c_{10} &=\frac{12(h^{10}-1)-11(h^4-1)(h^6-1)N}{23100 (h^2-1)^5 N^4}
\end{align}\end{subequations}
where we have again introduced $h = g+1$.
It is remarkable that $c_k$ coefficients with $k$ as high as 10 keep a quite 
tractable formulation. These formulas allow us to build a fast analytical 
approximation for $\mathscr{S}$, either as a normal distribution, or as a 
Gram-Charlier series.

\begin{figure}[htbp]
 \centering
 \subfigure[$N=2$]{\label{fig:GC_g2_N2}
\includegraphics[scale=0.265]{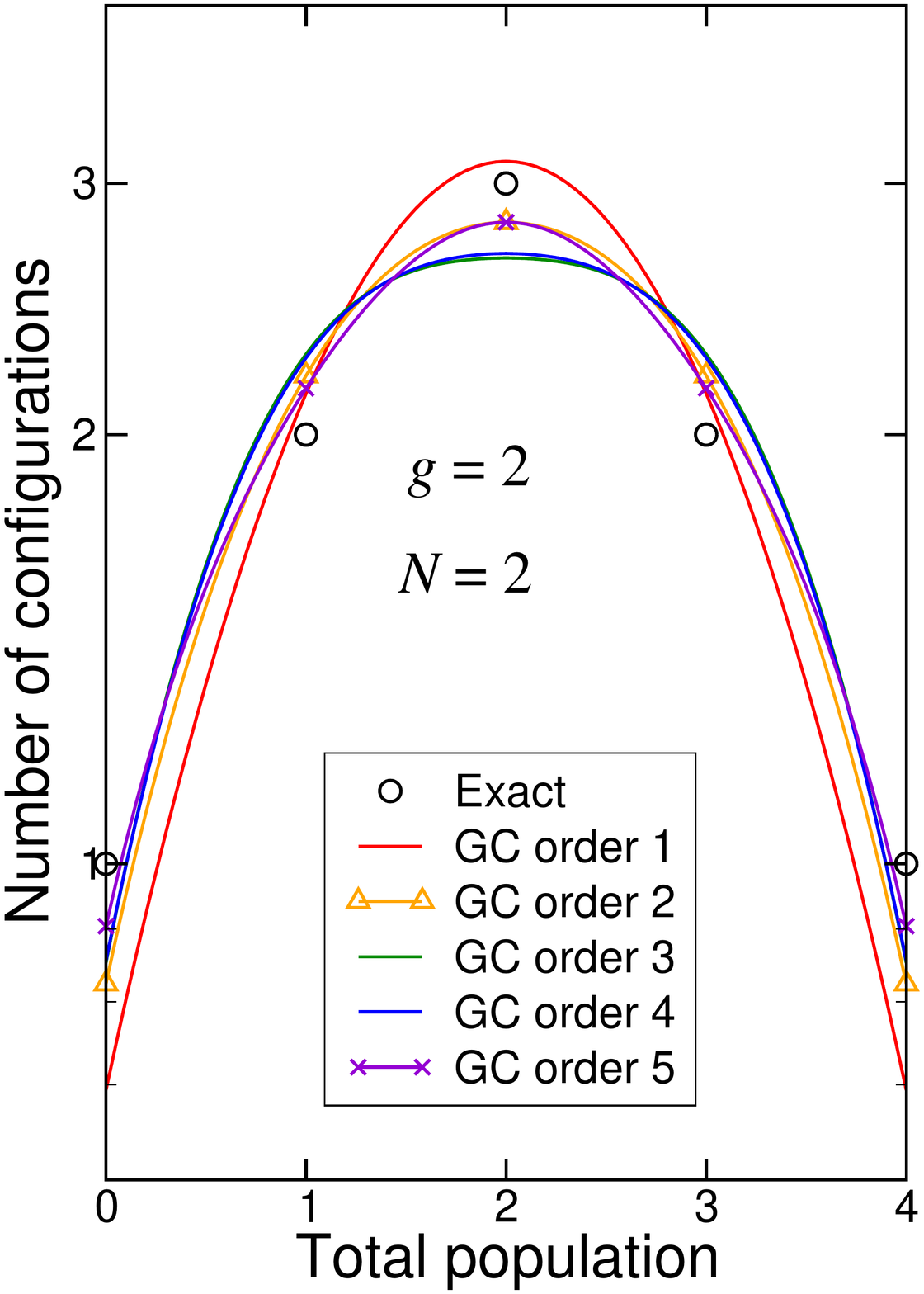}
}\goodgap%
\subfigure[$N=5$]{\label{fig:GC_g2_N5}
\includegraphics[scale=0.265]{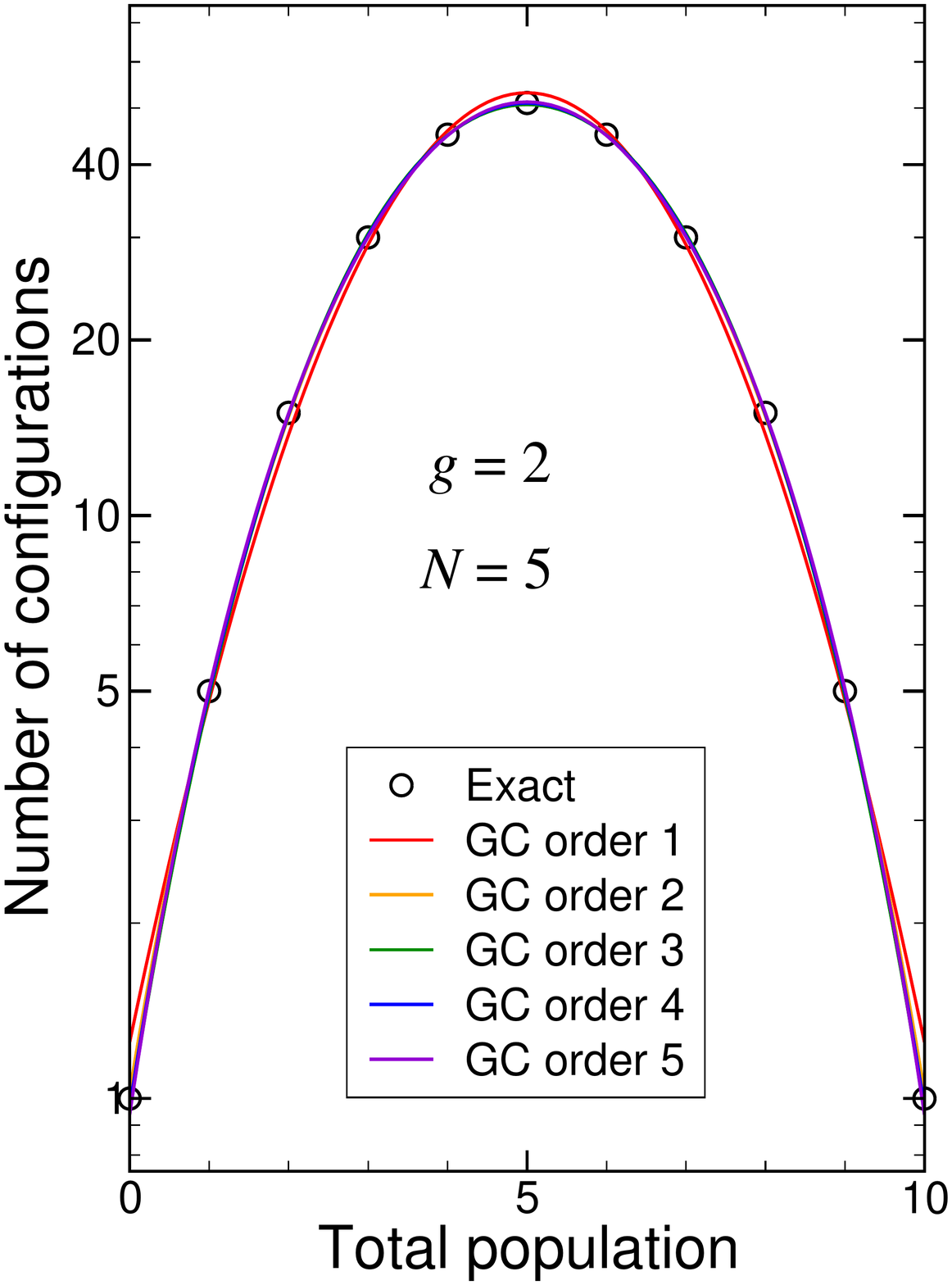}
}\goodgap%
\subfigure[$N=10$]{\label{fig:GC_g2_N10}
\includegraphics[scale=0.265]{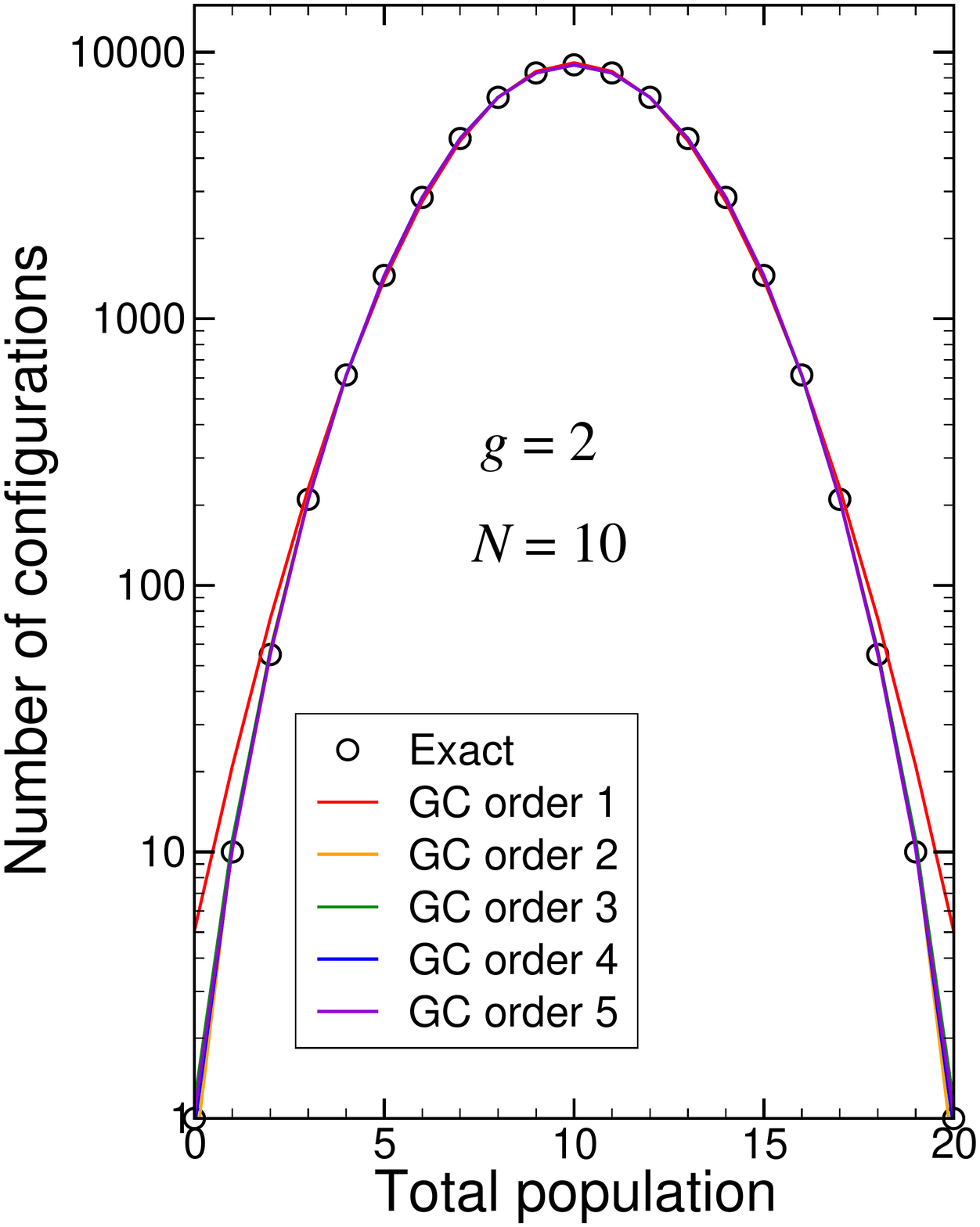}
}\\
 \caption{Comparison of the exact population distribution in $N$ subshells 
 of degeneracy $g=2$ with Gram-Charlier expansions at various orders. 
 The Gram-Charlier expansion is plotted as a continuous function of the total 
 population $Q$. In this figure, ``order $p$'' means that moments up to $k=2p$ 
 have been included in the expansion.\label{fig:testGC_g2}}
\end{figure}

\begin{figure}[htbp]
 \centering
 \subfigure[$N=2$]{\label{fig:testGC_g10_N2}
 \includegraphics[scale=0.260]{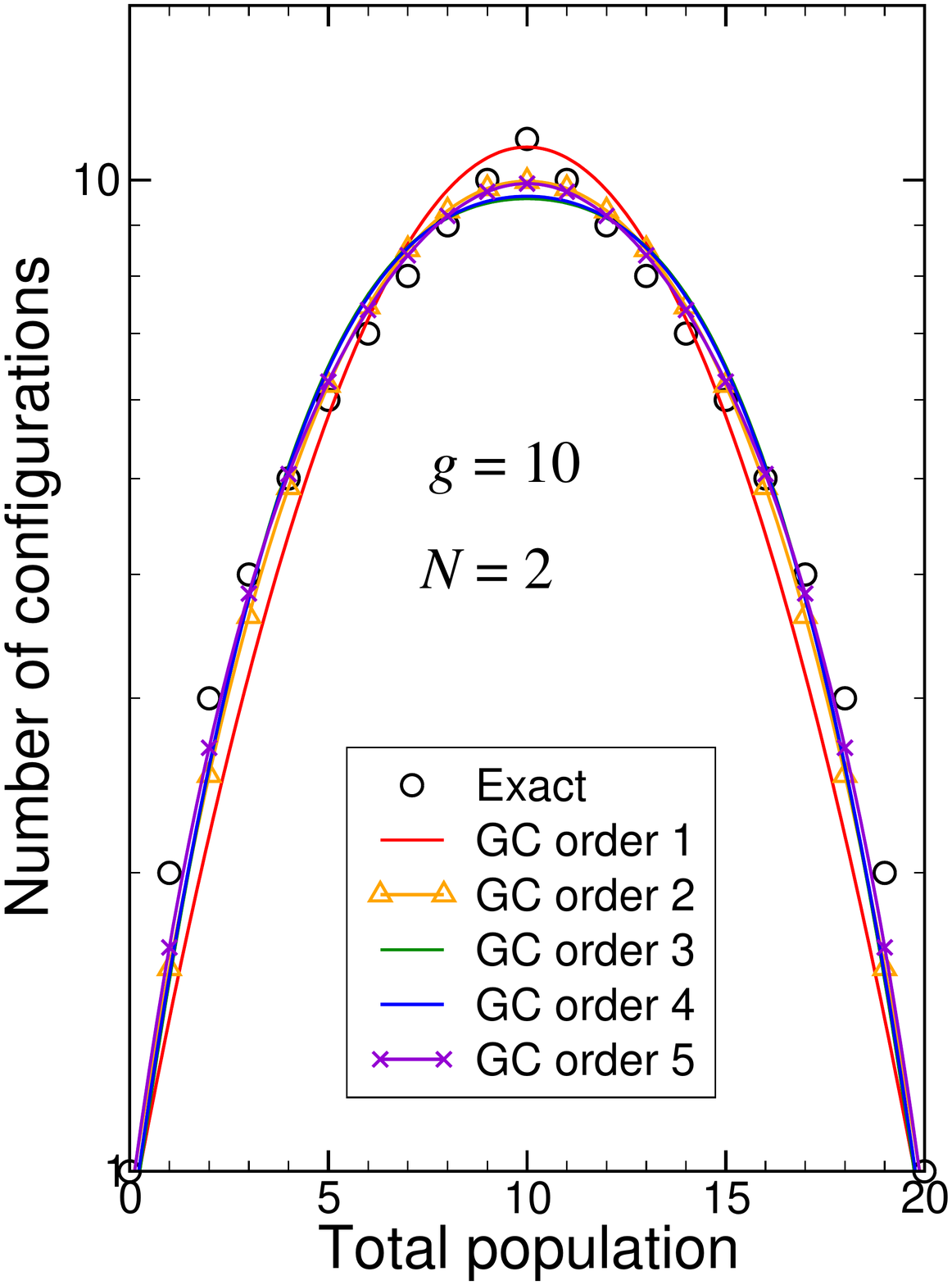}}
\goodgap%
\hspace{0.1cm}
\subfigure[$N=5$]{\label{fig:testGC_g10_N5}
\includegraphics[scale=0.260]{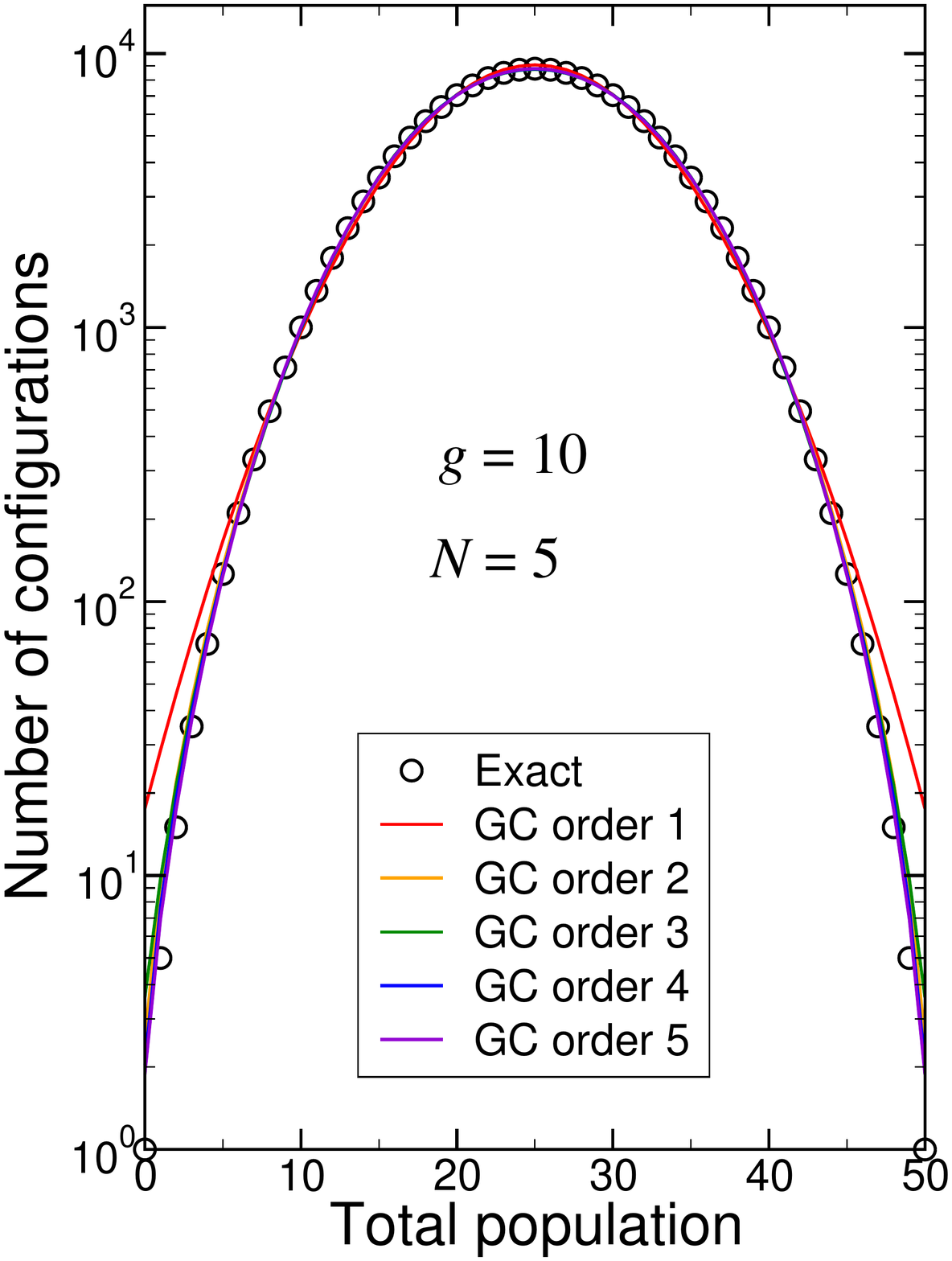}
}\goodgap\hspace{0.1cm}
\subfigure[$N=10$]{\label{fig:testGC_g10_N10}
\includegraphics[scale=0.260]{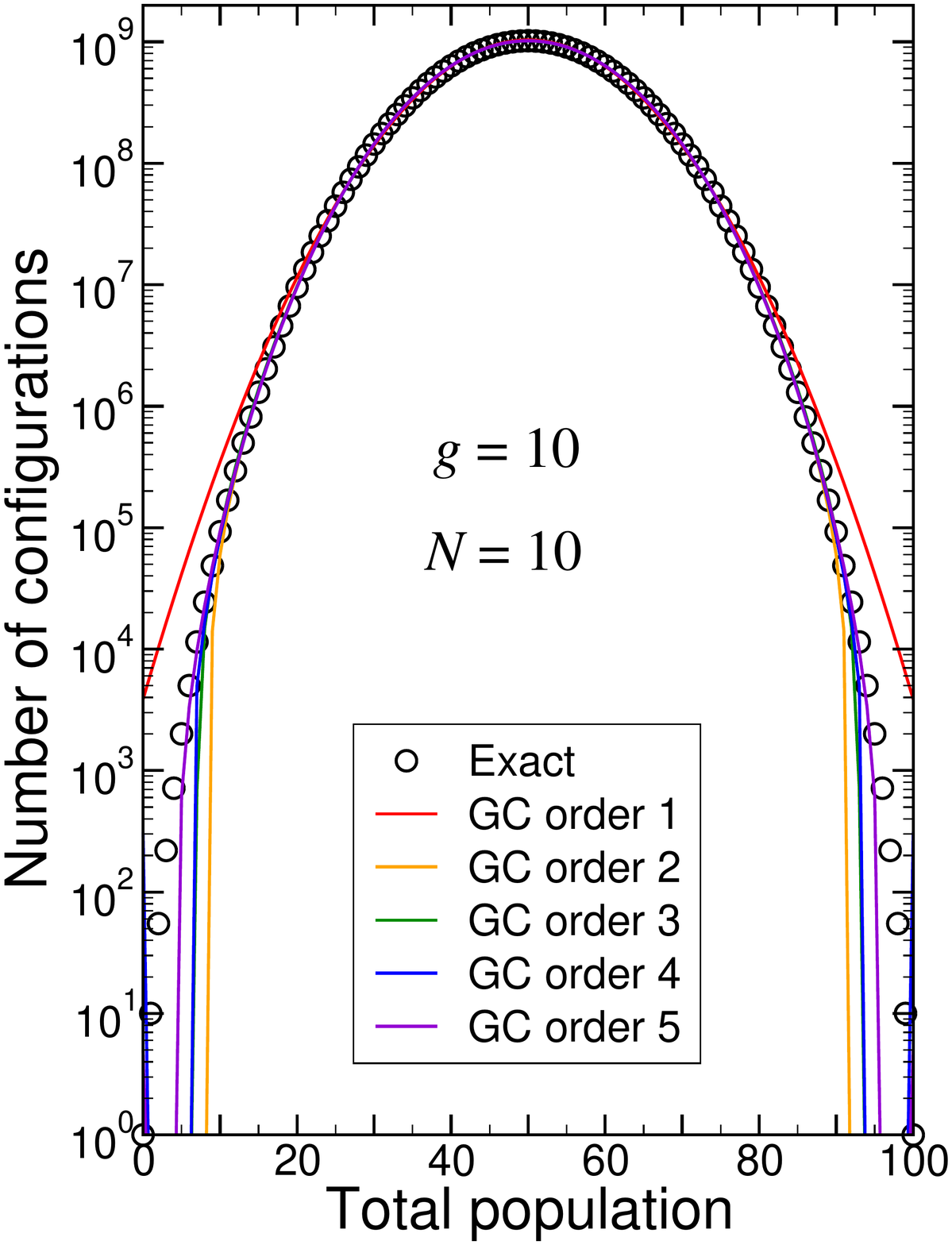}
}\\
 \caption{Comparison of the exact population distribution in $N$ 
 subshells of degeneracy $g=10$ with Gram-Charlier expansions at 
 various orders.}
 \label{fig:testGC_g10}
\end{figure}

Using the above relations (\ref{eq:Gram-Charlier},\ref{eq:CGC4-10}) we have 
compared the exact distribution $\mathscr{S}(g;N;Q)$ with Gram-Charlier 
expansions for several $(g,N)$ pairs on the whole $Q=0-g.N$ range of 
populations. Examples are given in Figs. \ref{fig:testGC_g2} and 
\ref{fig:testGC_g10} for $g=2$ and $g=10$ respectively. In each figure, 
cases $N=2,5$, and 10 have been studied. One observes that even the normal 
distribution, i.e., formula (\ref{eq:Gram-Charlier}) with all $c_k$ canceled, 
provides a reasonable approximation of the $\mathscr{S}(g;N;Q)$ value. 
Looking in more detail, in the wings of the distribution, the inclusion 
of at least the 2nd-order correction $c_4 He_4(X)$ in the Gram-Charlier 
expansion significantly improves the quality of the approximation. 
As mentioned above, the evaluation of such correction using the expression 
(\ref{eq:cGC4}) is straightforward.
One may notice a visible, though moderate, discrepancy in the case $N=2$, 
whatever the $g$ value. This may be easily understood by computing directly 
the $\mathscr{S}(g;N=2;Q)$ value. Using the recursion relations 
(\ref{eq:recurnbconfig}) and the initial value (\ref{eq:SgN1Q})
one may check that $\mathscr{S}(g;N;Q)$ expressed versus $Q$ are piecewise 
polynomials of degree $N-1$, with a unique definition on intervals of length 
$g$. Namely, one obtains 
\begin{equation}
 \mathscr{S}(g;2;Q)=g+1-\left|Q-g\right|\label{eq:S2Q}
\end{equation}
\begin{equation}
\mathscr{S}(g;3;Q)=\begin{cases}\frac12(Q+1)(Q+2)&\text{ if } 0\le Q\le g \\
 \frac12(g+1)(g+2)-(Q-g)(Q-2g) &\text{ if } g\le Q\le 2g \\
 \frac12(Q-3g-1)(Q-3g-2) &\text{ if } 2g\le Q\le 3g\end{cases}.
\end{equation}
Obviously, it quite difficult to approximate the triangle-shaped function 
(\ref{eq:S2Q}) with a normal distribution. The approximations at the various 
orders Gram-Charlier of $\mathscr{S}(2;2;Q)$ are given in table 
\ref{tab:GC_g2_N2}. 
It turns out that the maximum discrepancy is about 10 \%. 
For $Q=0$, the discrepancy decreases with the expansion order, while for 
$Q=1,2$ the first order is better than the next four orders. An optimum 
is reached at sixth order, and for higher orders the overall agreement 
deteriorates, with some oscillations. Finally, above 18th order, we have 
checked that the Gram-Charlier expansion clearly diverges.
These considerations concern the convergence analysis of the Gram-Charlier 
expansion more than the computational interest of this series, 
since for the lowest $N$ values, as seen in the above mentioned examples, 
simple piecewise polynomial expressions are available.

\begin{table}[thbp]
\begin{tabular}{cccccccccc}
\hline\hline
$Q$ & Exact & Order 1 & Order 2 & Order 3 & Order 4 & Order 5 & Order 6 & 
Order 7 & Order 8\\
\hline
0 & 1 & 0.694 & 0.824 & 0.852 & 0.855 & 0.904 & 0.991 & 1.070 & 1.114 \\
1 & 2 & 2.137 & 2.200 & 2.277 & 2.264 & 2.155 & 2.002 & 1.861 & 1.769 \\
2 & 3 & 3.109 & 2.818 & 2.660 & 2.679 & 2.818 & 3.003 & 3.174 & 3.294 \\
\hline\hline
\end{tabular}
\caption{Number of configurations as a function of the population $Q$ 
for $N=2$ subshells of degeneracy $g=2$: exact values and Gram-Charlier 
approximations. Order one is the normal distribution, order 2 includes 
the kurtosis contribution, etc.\label{tab:GC_g2_N2}}
\end{table}

As seen in figure \ref{fig:testGC_g10} dealing with a greater $g$ value, 
while the Gram-Charlier expansion at 2nd order (with the excess kurtosis 
accounted for) is quite acceptable in most of the $Q=0$ to $gN$ range, 
discrepancies are clearly visible for $Q\lesssim(gN)^{1/2}$, $Q\gtrsim 
gN-(gN)^{1/2}$. For such population values, the number of configurations 
$\mathscr{S}$ is usually orders of magnitude below its peak value 
$(g+1)^N/(2\pi\sigma^2)^{1/2}$, however one may be interested in 
approximations uniformly valid whatever $Q$. In this case it appears that the 
inclusion of more terms in the Gram-Charlier expansion improves its accuracy 
in the wings. Though this behavior is clear on subfigure 
\ref{fig:testGC_g10_N10}, we did not try to get a quantitative estimate of 
the Gram-Charlier order which provides a uniform approximation for the 
$\mathscr{S}(g=10;N=10;Q)$ values.

\subsection{Multiple-degeneracy case}
\begin{figure}[htbp]
 \centering
 \subfigure[$g_1=2,N_1=2$, $g_2=6,N_2=2$]{\label{fig:GC_s2p2}%
 \includegraphics[scale=0.32]{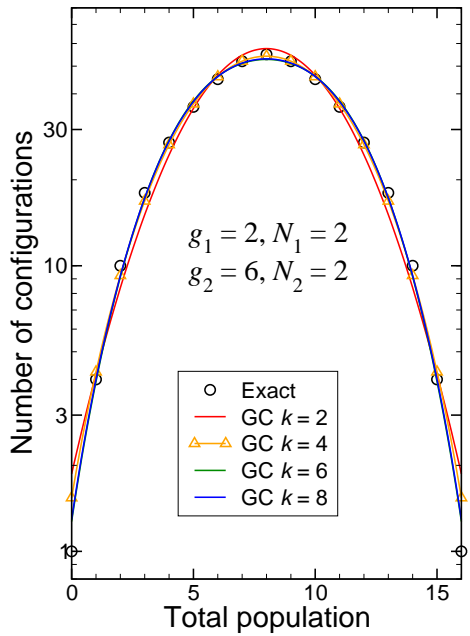}}
 \goodgap\hspace{0.1cm}
 \subfigure[$g_1=2,N_1=3$, $g_2=6,N_2=2$, $g_3=10,N_3=1$]%
 {\label{fig:GC_s3p2d1}%
 \includegraphics[scale=0.32]{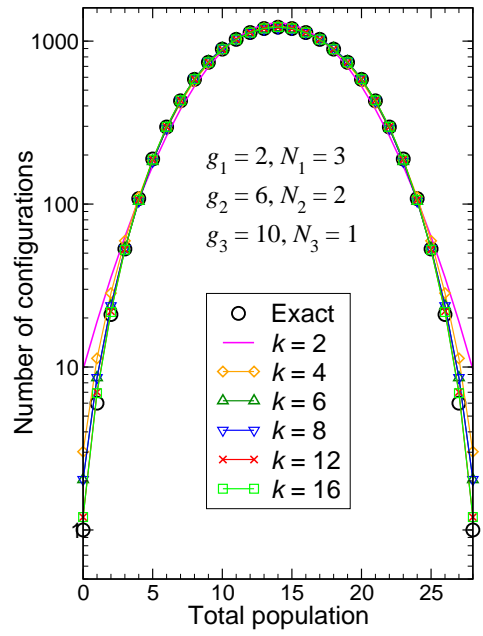}}
 \caption{Comparison of the exact population distribution with Gram-Charlier 
 expansions at various orders for the supershell s[2]p[2] (two subshells s 
 and two subshells p, for instance 2s2p3s3p) (a), and for the supershell 
 s[3]p[2]d[1] (for instance 1s2s2p3s3p3d) (b).\label{fig:testGC_multi}}
\end{figure}

Using the general expression (\ref{eq:cGC_vs_cumulants}) of the Gram-Charlier 
coefficients, and the cumulant value (\ref{eq:cumulant_gen}), one easily gets 
the first terms of the expansion
\begin{subequations}\begin{align}
c_4 &= -\frac{C_4}{20 C_2^2}\\
c_6 &= \frac{C_6}{105 C_2^3}\\
c_8 &= \frac{7 C_4^2-12 C_8}{5600 C_2^4}\\
c_{10} &= \frac{12 C_{10}-11 C_4C_6}{23100 C_2^5}
\end{align}\end{subequations}
which generalize the Eqs. (\ref{eq:CGC4-10}) in the multi-degeneracy case.
Such a procedure has been used first to analyze the population distribution 
in the case $t=2, g_1=2, N_1=2, g_2=6, N_2=2$, labeled s[2]p[2] for short. 
The Gram-Charlier analysis is presented in figure \ref{fig:GC_s2p2}. We note 
that, even though the number of subshells is small (4), the Gram-Charlier 
expansion with the first correction $c_4$ (orange curve and triangles) 
provides a fair approximation of the exact number. Moreover the Gram-Charlier 
formula, of statistical nature, would perform even better for more complex 
configurations with a greater number of subshells.

As a second example the Gram-Charlier approximation for the more complex 
supershell s[3]p[2]d[1] (for instance 1s2s2p3s3p3d) is analyzed on 
figure \ref{fig:GC_s3p2d1}. One checks that Gram-Charlier at second order 
($k=4$) is in fair agreement with the exact data. The 3rd order ($k=6$) 
improves again the agreement, with no significant gain at 4th order ($k=8$). 
The higher-order expansions $k=12, 16$ bring an improved agreement with the 
exact value, especially for the smallest and largest $Q$ values.

\begin{figure}[htbp]
\centering
\includegraphics[scale=0.4]{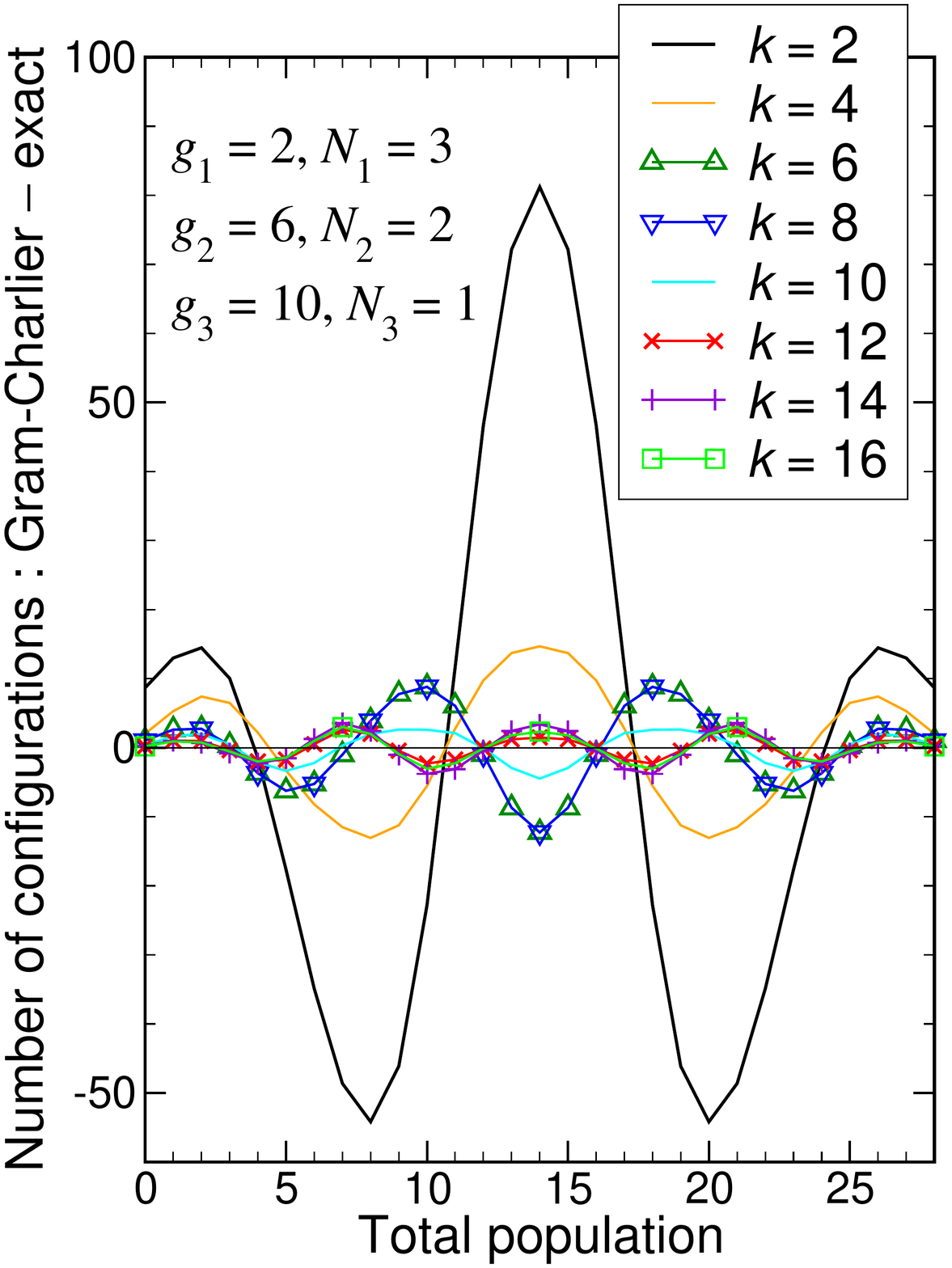}
\caption{Difference between the number of configurations obtained with the 
Gram-Charlier expansion truncated at various orders and the exact value for 
the supershell 1s2s2p3s3p3d. The $k=2$ curve corresponds to the normal 
distribution, $k=4$ is the Gram-Charlier series involving up to the $He_4$ 
polynomial or second-order approximation, etc. Though exact values are only 
defined for integer populations $Q$, lines are drawn as a visual guide.
\label{fig:GC-exact_s3p2d1}}
\end{figure}

As a rule one may check that the accuracy of the Gram-Charlier expansion 
globally increases with the order, though some oscillations are noticed. 
As an example, in figure \ref{fig:GC-exact_s3p2d1} we have plotted the 
difference between the Gram-Charlier approximation (\ref{eq:Gram-Charlier}) 
truncated at various orders and the exact number of configurations. In this 
particular case, a good compromise between the quality of the expansion and 
the computational cost is reached for $k=10$, i.e., with five terms in the 
sum.
As shown below, a more complete numerical analysis involving higher orders 
demonstrates that the Gram-Charlier series is indeed divergent.

\section{Analysis of population distribution using Edgeworth expansion}
\label{sec:Edgeworth}
It has been mentioned that some distributions get a better representation 
in terms of Edgeworth series rather than of Gram-Charlier series 
\cite{Blinnikov1998}. Another interest of the Edgeworth expansion is that it 
is directly expressed in terms of cumulants rather than of centred moments. 
The Edgeworth series is an expansion versus powers of the standard deviation 
$\sigma$, defined as
\begin{subequations}\begin{gather}
 E(Q) = G\frac{\exp(-x^2/2)}{\sqrt{2\pi}\sigma} 
  \left\{ 1+\sum_{s=1}^\infty\sigma^s\sum_{\{k_m\}}He_{s+2r}(x)
  \prod_{m=1}^s\frac{1}{k_m!}\left(\frac{S_{m+2}}{(m+2)!}\right)^{k_m} 
  \right\}\label{eq:Edgeworth}\\
  \text{with\quad}S_n=\kappa_n/\sigma^{2n-2}\text{,\quad}r=k_1+k_2+\cdots k_s
\end{gather}
$x$ being the reduced variable
\begin{equation}x=(Q-\left<Q\right>)/\sigma\end{equation}
and where the index $\{k_m\}$ refer to all $s$-uple indices verifying 
\begin{equation}k_1+2k_2+\cdots+sk_s=s.\end{equation}
\end{subequations}
As for Gram-Charlier expansion, this series involves only even $s$ orders. 
The sum over $s$ is replaced by a finite sum up to some $s_\text{trunc}$, 
which is chosen as discussed below.

\begin{figure}[htbp]
\begin{center}
\includegraphics[scale=0.6,angle=0]{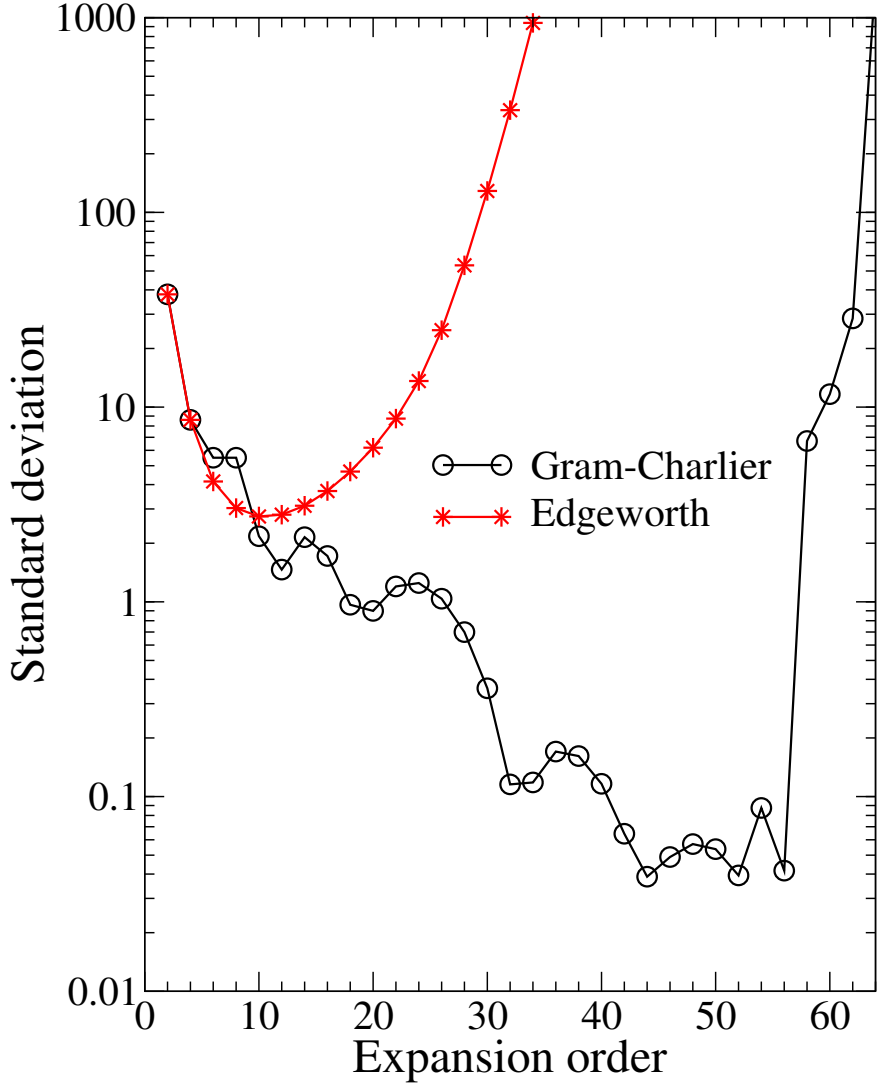}
\caption{Standard mean deviation $\left[\sum_{Q=0}^{Q_\text{max}}
\left(N_\text{app}(Q;s_\text{trunc})-N_\text{exact}(Q)\right)^2/
(Q_\text{max}+1)\right]^{1/2}$ 
for the number of configurations $Q$ computed exactly or using expansions 
truncated at various orders $s_\text{trunc}$. Configurations are generated 
from the $1s2s2p3s3p3d$ supershell and approximations are those obtained from 
Gram-Charlier and Edgeworth series. Only even $s_\text{trunc}$ values are 
plotted since odd-order terms in the expansions vanish.
\label{fig:GC_vs_Edgeworth_s3p2d1}}
\end{center}
\end{figure}

In order to compare Edgeworth and Gram-Charlier expansions we have plotted 
in figure \ref{fig:GC_vs_Edgeworth_s3p2d1} the average deviation 
\begin{equation}
\Delta_\text{app}(s_\text{trunc}) = \left[\sum_{Q=0}^{Q_\text{max}}
 \left(N_\text{app}(Q;s_\text{trunc})-N_\text{exact}(Q)\right)^2/
 (Q_\text{max}+1)\right]^{1/2}
\end{equation}
for the $1s2s2p3s3p3d$ supershell as a function of $s_\text{trunc}$. In the 
above formula $Q_\text{max}$ is the maximum occupation number of the 
supershell $\sum_i g_iN_i$, 28 in the present case, 
$N_\text{app}(Q;s_\text{trunc})$ is the approximate number of configurations 
with occupation $Q$ computed with Gram-Charlier (\ref{eq:Gram-Charlier}) or 
Edgeworth (\ref{eq:Edgeworth}) truncated series. 
A truncation order $s_\text{trunc}=2$ corresponds to 
the normal distribution, the truncation $s_\text{trunc}=4$ corresponds to 
terms involving the Chebyshev-Hermite polynomial $He_4(X)$, etc. On this 
graph, it appears that both expansions provide an acceptable representation 
of the number of configurations for the low values of $s_\text{trunc}$. 
Truncating the expansion at $s_\text{trunc}=10$, i.e., keeping four 
correction terms to the normal distribution, provides the best 
approximation in case of Edgeworth series. In this case, the relative error 
$\left[\sum_{Q=0}^{Q_\text{max}}\left(N_\text{app}(Q;s_\text{trunc})/
N_\text{exact}(Q)-1\right)^2/(Q_\text{max}+1)\right]^{1/2} \simeq 0.4$ for 
Edgeworth expansion, while the absolute deviation plotted on 
figure~\ref{fig:GC_vs_Edgeworth_s3p2d1} is 2.74. This apparently poor agreement 
is due to the large error in the $Q=0$ approximate value : $N_\text{Edgeworth}
(Q=0;s_\text{trunc}=10)\simeq-0.519$ while $N_\text{exact}(Q=0)=1$. 
However large values around $Q_\text{max}/2$ are better represented : 
indeed one has $N_\text{Edgeworth}(Q=14;s_\text{trunc}=10)\simeq1221.79, 
N_\text{exact}(Q=14) = 1217$. The general behavior is quite different for 
$s_\text{trunc}$ above 10: while Gram-Charlier accuracy still improves with 
$s_\text{trunc}$, the Edgeworth-expansion accuracy deteriorates rapidly. 
As seen on the graph, for very large values ($s_\text{trunc}>56$), 
the Gram-Charlier expansion also diverges rapidly. 
This behavior has been mentioned previously \cite{Blinnikov1998}, 
but our conclusion is that Gram-Charlier expansion provides here a better 
approximation than Edgeworth expansion. Our conclusion is also at variance 
with the observation by de Kock \textit{et al.} \cite{deKock2011} who claim 
that Edgeworth series strongly outperforms Gram-Charlier series. In our 
opinion this difference comes from the fact that we are dealing here with a 
\emph{discrete} distribution, defined only for integer values, and that this 
distribution is not an analytical function of $Q$ but a piecewise polynomial.

\section{Conclusion}\label{sec:concl}

We found three explicit formulas for the number of atomic configurations. Although the best way to compute such a quantity remains probably the double recurrence on the numbers of electrons and orbitals, the new expressions may be of interest in order to get new relations for the number of atomic configurations, using the numerous properties, identities and sum rules for binomial and multinomial coefficients.
Using a two-variable generating function, we have derived several recurrence 
relations, not published before up to our knowledge. Using the same 
generating function, the moments of the distribution have received an 
analytical expression. It allowed us to provide explicit expressions for 
moments up to the twelfth, though higher-order moments could be obtained too. 
The case of multiple value for the subshell degeneracy has been addressed 
using the cumulant formalism. We have shown that the cumulants receive a very 
simple expression whatever the order. This allowed us to obtain centred 
moments explicitly for $k$ up to 12. 
A Gram-Charlier analysis has shown that an expansion with two terms is in 
acceptable if not fair agreement with the exact number of configurations, 
though the series is not convergent. We have found that the Edgeworth expansion 
provides an equivalent accuracy if few terms are kept, though it diverges 
much more rapidly than the Gram-Charlier series.

\appendix
\section{Numbering the partitions defined by subset populations}\label{sec:numparts}
The purpose of this appendix is to enumerate the partitions of $n$ distinct objects 
knowing that there are $n_1$ subsets of population 1, $n_2$ subsets of population 2, 
\dots $n_k$ subsets of population $k$. In the main text one has $n=k$ though this 
constraint is not required for the present derivation. Conversely one must have 
\begin{equation}n=n_1+2n_2+\cdots kn_k.\end{equation}
The generation of these partitions may be done in $k+1$ steps. In the first step, 
one selects the $n_1$ elements in single-element subsets, the $2n_2$ elements in 
twofold subsets, up to the $kn_k$ elements in the subsets of population $n_k$. 
The number of possibilities at this step is
\begin{equation}\label{eq:partition_p0}
p_0=\binom{n}{n_1}\binom{n-n_1}{2n_2}\dots
 \binom{n-n_1\cdots-(k-2)n_{k-2}}{(k-1)n_{k-1}}
 \binom{n-n_1\cdots-(k-1)n_{k-1}}{kn_k}
 = \frac{n!}{\prod_{j=1}^k (j n_j)!}.
\end{equation}
At the next $k$ steps one must choose, for any $j$ from 1 to $k$, how to partition 
$jn_j$ objects in $n_j$ subsets. This operation is performed by first selecting 
$j$ objects among $jn_j$, then $j$ more objects among $j(n_j-1)$, i.e., 
repeating the selection process $n_j-1$ times. When this multiple selection is 
completed, one gets $n_j!$ identical solutions, since the order of the subsets 
is not significant. Therefore the number of possibilities at step $j$ is
\begin{equation}\label{eq:partition_pj}
p_j=\frac{1}{n_j!}\binom{jn_j}{j}\binom{(j-1)n_j}{j}\cdots
 \binom{2j}{j}\binom{j}{j}=\frac{1}{n_j!}\frac{(jn_j)!}{(j!)^{n_j}}.
\end{equation}
Multiplying $p_0$ given by Eq. (\ref{eq:partition_p0}) by the product of $p_j$'s 
provided by Eq. (\ref{eq:partition_pj}) one gets the desired number of partitions
\begin{equation}\label{eq:partition_number}
\mathscr{P}(n;n_1,n_2\cdots,n_k)=\frac{n!}{\prod_{j=1}^k n_j!(j!)^{n_j}}.
\end{equation}

\section{Coefficients of the Gram-Charlier expansion as  a function of the 
cumulants}\label{sec:app_GC_vs_cumul}
The generating function of the cumulants is defined as
\begin{equation}\mathscr{K}(t)=\sum_{n=1}^\infty \kappa_n\frac{t^n}{n!}
 =\log\left(\left<\exp(tQ)\right>\right).
\end{equation}
In the case of the Gram-Charlier expansion the integral 
$\left<\exp(tQ)\right>$ is easily obtained as 
\begin{equation}e^{\mathscr{K}(t)} = \left<\exp(tQ)\right> = 
 \int dQ \frac{\exp(tQ-Q^2/2\sigma^2)}{\sqrt{2\pi\sigma^2}}
 \left[1+\sum_{n>2} c_n He_n(Q/\sigma)\right].
\end{equation}
Using the Rodrigues formula for $He_n(X)$ and repeated integration by parts 
one easily gets 
\begin{equation}
\int_{-\infty}^{+\infty}dQ \:e^{tQ-X^2/2\sigma^2}He_n(Q/\sigma) = 
 (\sigma t)^n\exp(\sigma^2t^2/2)
\end{equation}
from which one has the average over Gram-Charlier distribution
\begin{equation}\label{eq:moyexpQTdistrGC}
 \left<\exp(tQ)\right> = \int_{-\infty}^{+\infty}dQ\;e^{tQ-X^2/2\sigma^2} 
  \left[1+\sum_{n>2}c_n He_n(Q/\sigma)\right] =
 e^{\sigma^2t^2/2}\left[1+\sum_{n>2}(\sigma t)^nc_n\right].
\end{equation}
The exponential of the generating function of cumulants is, for any 
\emph{centred} distribution (i.e., such as $\kappa_1=0$),
\begin{equation}
e^{\mathscr{K}(t)} = \exp\left(\sum_{n=1}^\infty\kappa_n\frac{t^n}{n!}\right)
 = e^{\kappa_2t^2/2}\exp\left(\sum_{n=3}^\infty\kappa_n\frac{t^n}{n!}\right).
\end{equation}
Identifying this expression with the average (\ref{eq:moyexpQTdistrGC}), one 
writes
\begin{subequations}\begin{equation}
 1+\sum_{n\ge3}(\sigma t)^nc_n 
 =\exp\left(\sum_{n=3}^\infty\kappa_n\frac{t^n}{n!}\right)
 =\exp\left(\sum_{n=1}^\infty x_n\frac{t^n}{n!}\right)
 =\sum_{m=0}^\infty\frac1{m!}
  \left(\sum_{n=1}^\infty x_n\frac{t^n}{n!}\right)^m\label{eq:som_xn_m}
\end{equation}
where we have defined
\begin{equation}\label{eq:defx}
 x_1=0\text{,\quad}x_2=0\text{,\quad}x_n=\kappa_n\text{ if }n\ge3.
\end{equation}\end{subequations}
The $m$th power in the sum (\ref{eq:som_xn_m}) may be computed with the 
identity (see section 24.1.2 in Ref. \cite{Abramowitz1972})
\begin{subequations}\begin{equation}\label{eq:expr_fn_gen_cumul}
\left(\sum_{n=1}^\infty x_n\frac{t^n}{n!}\right)^m = 
 m!\sum_{n=m}^\infty\frac{t^n}{n!}\sum_{a_1,a_2,\cdots a_n}
 \mathscr{P}(n;a_1,a_2\cdots,a_n)\:x_1^{a_1}x_2^{a_2}\cdots x_n^{a_n}
\end{equation}
with the above definition (\ref{eq:partition_number}) of the partition 
number $\mathscr{P}$, and where integer indices $a_1,a_2,\cdots a_n$ are 
constrained by
\label{eq:sum_ai}\begin{gather}
a_1+a_2+\cdots+a_n=m\\
a_1+2a_2+\cdots+na_n=n.\label{eq:sum_iai}
\end{gather}\end{subequations}
Identifying terms in $t^n$ in Eqs. (\ref{eq:som_xn_m},
\ref{eq:expr_fn_gen_cumul}), one has 
\begin{equation}\label{eq:cn_sum_m_ai}
\sigma^n c_n = \frac{1}{n!}\sum_{m\le n}\sum_{a_1,a_2,\cdots a_n}
 \mathscr{P}(n;a_1,a_2\cdots,a_n)\:x_1^{a_1}x_2^{a_2}\cdots x_n^{a_n}
\end{equation}
where the sum on $a_i$ follows the constraints (\ref{eq:sum_ai}). 
One will note that, since the $a_i$ are nonnegative, one has 
\begin{equation}
 m=a_1+a_2+\cdots+a_n\le a_1+2a_2+\cdots+na_n=n,
\end{equation}
therefore in the multiple sum (\ref{eq:cn_sum_m_ai}) one may ignore the sum 
over $m$, since this index is only intended to collect terms in the sum. 
One has then
\begin{equation}\label{eq:c_n_sum_ai}
\sigma^n c_n = \frac{1}{n!}\sum_{a_1,a_2,\cdots a_n}
 \mathscr{P}(n;a_1,a_2\cdots,a_n)\:x_1^{a_1}x_2^{a_2}\cdots x_n^{a_n}
\end{equation}
where only the second constraint (\ref{eq:sum_iai}) has been kept. Accounting 
for $x_i$ definitions (\ref{eq:defx}), one notes that only terms with 
$a_1=0, a_2=0$ contribute and one gets the Gram-Charlier-series coefficient 
\begin{subequations}\begin{align}\label{eq:cGC_vs_cumulants}
 c_n &= \frac{1}{\sigma^n n!}
  \sum_{\substack{a_3,\cdots a_n\\3a_3+\cdots+na_n=n}}
 \mathscr{P}(n;0,0,a_3\cdots,a_n)\:\kappa_3^{a_3}\cdots\kappa_n^{a_n}\\
 &= \sum_{\substack{a_3,\cdots a_n\\3a_3+\cdots+na_n=n}}
  \frac{1}{a_3!}\left(\frac{\kappa_3}{3!\sigma^3}\right)^{a_3}\cdots
  \frac{1}{a_n!}\left(\frac{\kappa_n}{n!\sigma^n}\right)^{a_n}.
\end{align}\end{subequations}


\bibliography{jrnlabbr,nconf}
\bibliographystyle{plain}

\end{document}